\documentclass[traditabstract]{aa}

\usepackage{natbib}
\usepackage{amssymb,verbatim}\usepackage{graphicx}
\usepackage{hyperref}
\newcommand{\mincir}{\raise -2.truept\hbox{\rlap{\hbox{$\sim$}}\raise5.truept
\hbox{$<$}\ }}
\newcommand{\magcir}{\raise -2.truept\hbox{\rlap{\hbox{$\sim$}}\raise5.truept
\hbox{$>$}\ }}
\newcommand{\siml}{\raise -2.truept\hbox{\rlap{\hbox{$\sim$}}\raise5.truept
\hbox{$<$}\ }}
\newcommand{\simg}{\raise -2.truept\hbox{\rlap{\hbox{$\sim$}}\raise5.truept
\hbox{$>$}\ }}
\newcommand{\be}{\begin{equation}}
\newcommand{\ee}{\end{equation}}
\newcommand{\ba}{\begin{eqnarray}}
\newcommand{\ea}{\end{eqnarray}}

\newcommand {\h} {$h_{70}^{-1}$ Mpc$\;$}

\newcommand {\ks} {km~s$^{-1} \;$}
\newcommand {\kss} {km~s$^{-1}$}

\newcommand {\mqua} {$\times 10^{14}\;h_{70}^{-1}\;M_{\odot} \;$}

\newcommand{\degree}{\ensuremath{\mathrm{^\circ}}}

\newcommand{\arcs}{\ensuremath{\arcmm\hskip -0.1em\arcmm \;}}
\newcommand{\arcmm}{\ensuremath{\mathrm{^\prime}}}

\newcommand{\dotsec}{\,\rlap{\hbox{$\mathrm{^s}$}}{\hbox{$.$}}\,}

\begin{document}

\titlerunning{A textbook example of ram-pressure stripping}

\title{A textbook example of ram-pressure stripping in  \\
the Hydra A/A780 cluster}

\author{S. De Grandi\inst{1}, D. Eckert\inst{2,3} \and S. Molendi\inst{3} \and M. Girardi\inst{4,5}  \and E. Roediger\inst{6} \and M. Gaspari\inst{7} \and F. Gastaldello\inst{3}, S.  Ghizzardi \inst{3},  M. Nonino\inst{4} \and M. Rossetti\inst{3,8}  }
\institute{
INAF - Osservatorio Astronomico di Brera, via E. Bianchi 46, 23807 Merate, Italy\\
\email{sabrina.degrandi@brera.inaf.it}
\and
Department of Astronomy, University of Geneva, ch. d'Ecogia 16, 1290 Versoix, Switzerland
\and
INAF - IASF-Milano, Via E. Bassini 15, 20133 Milano, Italy
\and
Dipartimento di Fisica dell'Universit\`a degli Studi di Trieste - Sez. di Astronomia, via Tiepolo 11, 34143 Trieste, Italy
\and
INAF - Osservatorio Astronomico di Trieste, via Tiepolo 11, 34143 Trieste, Italy
\and
E.A. Milne Centre for Astrophysics Department of Physics \& Mathematics University of Hull, Hull, HU6 7RX, United Kingdom
\and
Department of Astrophysical Sciences, Princeton University, Princeton, NJ 08544, USA; Einstein and Spitzer Fellow
\and
Universit\`a degli Studi Milano, via Celoria 16, 20133 Milano, Italy
}

\abstract{In the current epoch, one of the main mechanisms driving the growth of galaxy clusters is the continuous accretion of group-scale halos. In this process, the ram pressure applied by the hot intracluster medium on the gas content of the infalling group is responsible for stripping the gas from its dark-matter halo, which gradually leads to the virialization of the infalling gas in the potential well of the main cluster. Using deep wide-field observations of the poor cluster Hydra A/A780 with \emph{XMM-Newton} and \emph{Suzaku}, we report the discovery of an infalling galaxy group 1.1 Mpc south of the cluster core. The presence of a substructure is confirmed by a dynamical study of the galaxies in this region. A wake of stripped gas is trailing behind the group over a projected scale of 760 kpc. The temperature of the gas along the wake is constant at kT $\sim1.3$ keV, which is about a factor of two less than the temperature of the surrounding plasma. We observe a cold front pointing westwards compared to the peak of the group, which indicates that the group is currently not moving in the direction of the main cluster, but is moving along an almost circular orbit. The overall morphology of the group bears remarkable similarities with high-resolution numerical simulations of such structures, which greatly strengthens our understanding of the ram-pressure stripping process.}

\maketitle

\section{Introduction}

Gravitationally-bound structures in the Universe are thought to grow hierarchically through the merging and accretion of smaller structures throughout cosmic time, until they form the most massive galaxy clusters we see today \citep[e.g.,][]{springel06}. In the outer regions of galaxy clusters, we can observe the accretion of galaxies and groups of galaxies onto the main dark-matter halo, and they contribute significantly to the growth of galaxy clusters in mass, member galaxies and hot gas \citep[e.g.,][]{berrier09,genel10,delucia12}. While the merging processes of two or more massive entities are well-studied \citep[and references therein]{markevitch07_rev,owers09}, the processes leading to the gentle accretion of smaller halos (group scale) onto massive clusters, which represent the main channel \citep[$\sim 70\%$,][]{berrier09} of cosmic structure growth, have been more elusive. During infall, the ram pressure applied by the ambient intracluster medium (ICM) is responsible for stripping the gas from its original halo and heating it up \citep[e.g.][]{gunn72,kawata08,mccarthy08b}, leading to the virialization of the gas in the main dark-matter halo.

This picture is confirmed by observations of X-ray \citep[e.g.][]{sun05,machacek06,sun07,randall08,zhang13} and atomic (HI) and molecular galaxy wakes behind individual infalling galaxies in the nearest galaxy clusters \citep[e.g.][]{boselli14,abramson11,sun07,chung07}. Observations of such features are of prime importance for our understanding of the mechanisms leading to the build-up of galaxy clusters and provide a striking confirmation of the hierarchical scenario of structure formation. The $\Lambda$CDM paradigm predicts that at the present epoch there should be approximately one accreting group with a gravitational mass in the range of a few $10^{13}M_\odot$ per massive cluster and per Gyr \citep{dolag09}. There are a few examples of accreting groups directly observed in X-rays such as A2142 \citep{eckert14}, NGC 4389 in Coma \citep{neumann03} and the southern group in A85 \citep{kempner02,durret05,ichinohe15}.

In addition to the stripping properties, these features are valuable for the study of ICM physics. Indeed, the survival of the stripped gas in contact with the surrounding ICM can tell us about the conduction timescale in the medium, which is usually found to be much longer than expected from pure Spitzer conductivity \citep{eckert14,sanders13_sci}. Additionally, \cite{roediger15a,roediger15b} used high-resolution simulations to study the properties of infalling galaxies, varying the viscosity of the ICM plasma. It was found that the morphology of the tails strongly depends on the viscosity of the fluid: while in the inviscid case Kelvin-Helmholtz (KH) instabilities rapidly develop and induce a fast mixing of the plasma, a high viscosity suppresses KH instabilities, which results in long, X-ray bright tails.

In 2012 we started an \emph{XMM-Newton} program to look for infalling gas clumps in two nearby clusters, A2142 and Hydra A \citep[A780,][]{abell89}. This program allowed us to discover a striking example of infalling substructures in A2142, namely we discovered in this cluster a galaxy group with a mass of a few $10^{13}$ M$_\odot$ that appears to be almost completely stripped from its hot gas \citep{eckert14}, which constitutes a spectacular tail extending over 800 kpc (the longest ram-pressure-stripped tail observed so far).
In this paper we focus our study on the outskirts of the other observed cluster, Hydra A, where we discovered another accreting galaxy group, centered on the galaxy LEDA 87445 \citep{smith04} $\sim 17^{\prime}$ south-east from the center of Hydra A cluster.
This infalling group shows a long low surface brightness tail (up to 760 kpc) and a density contact discontinuity (merger cold front), in the opposite direction from the tail, that give important indications on its motion and ram-pressure stripping properties.
We complement the \emph{XMM-Newton} data with a deep \emph{Suzaku} observation of Hydra A and optical data from the literature.

This paper is structured as follows: in Sect. \ref{sec:data} we describe our {\it XMM-Newton} (Sect. \ref{sec:xmm}) and {\it Suzaku} (Sect. \ref{sec:suzaku}) data sets together with their respective data reduction, imaging and spectral analysis techniques. In Sect. \ref{sec:results}, we describe the results of the data analysis, concentrating on the description of the morphology (Sect. \ref{sec:morph}) and spectral properties of the new X-ray infalling group (Sect. \ref{sec:specreg}). In Sect. \ref{sec:dynamics}, we analyze the available optical data from the literature for the Hydra A cluster galaxies and quantify the dynamical properties of the group.
In Sect. \ref{sec:disc}, we interpret and discuss our results:
thermodynamic and ram-pressure stripping properties of the gas removed from the group are derived in Sect. \ref{sec:profx}, \ref{sec:north} and \ref{sec:ram}. In Sect. \ref{sec:motion} we estimate the orbit of the group and in Sect.\ref{sec:simul} we compare our findings with recent numerical simulations. Our main results are summarize in Sect. \ref{sec:concl}.

Throughout the paper, we assume a $\Lambda$CDM cosmology with $H_o=70$ km s$^{-1}$, $\Omega_m=0.3$ and $\Omega_\Lambda=0.7$. At the redshift of Hydra A ($z=0.0539$), this corresponds to $1^{\prime\prime}=1.05$ kpc. The average temperature of this cluster is $\sim 3.5$ keV \citep{david01}; \cite{sato12} derived R500\footnote{For a given overdensity $\Delta$, $R_\Delta$ is the radius for which $M_\Delta/(4/3\pi R_\Delta^3)=\Delta\rho_c$} $= 811\pm10$ kpc ($\sim 13$ arcmin) and R200 $= 1189\pm96$ kpc ($\sim 20$ arcmin) from the fitting of the Hydra A hydrostatic mass with the NFW universal mass profile. All the quoted errors hereafter are at the $1\sigma$ level.

\begin{figure}
\centerline{\resizebox{\hsize}{!}{\includegraphics[angle=0]{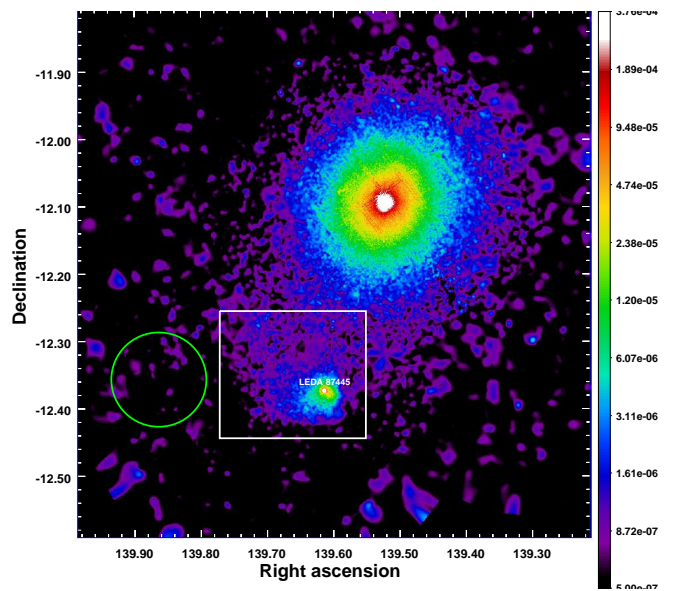}}}
\caption{Adaptively-smoothed, vignetting-corrected XMM/EPIC mosaic image of the Hydra A cluster in the [$0.7-1.2$] keV band. The group located around the galaxy LEDA 87445 is shown by the white square.  The green circle shows the region used to estimate the local background components.}
\label{fig:mosaic_xmm}
\end{figure}

\section{Data analysis}
\label{sec:data}

\subsection{\emph{XMM-Newton}}
\label{sec:xmm}

\subsubsection{Data reduction}
\label{sec:datared}

Hydra A / Abell 780 (z = 0.0539) is a cool-core cluster with a virial temperature of $\sim 3.5$ keV and a total flux of $2.7\times 10^{-11}$ ergs cm$^{-2}$ s$^{-1 }$ ($0.5-2.0$ keV band).  Except for cavities in the central regions caused by a giant AGN outburst \citep{mcnamara00}, its X-ray morphology is relaxed.  Hydra A was extensively observed with \emph{XMM-Newton} both on-axis (for a total nominal exposure time of $\sim 160$ ks, Obs. ID 0109980301-501, 0504260101) and off-axis
(6 pointings along different directions, $\sim 280$ ks total; PI: Eckert, Obs. ID. 0694440301-401-701-801, 0725240101, 0761550101)
thus we obtained a mosaic of this cluster covering the entire azimuth out to the virial radius.
We processed  the observations using the ESAS tasks as provided in SAS v.13.5. After soft protons cleaning procedure (with the {\it mos-filter} and {\it pn-filter} ESAS tasks, \citealt{snowden08}) the total available clean exposure time for the south region of Hydra A is 82 ks for MOS1, 88 ks for MOS2 and 27 ks for pn.

\subsubsection{Imaging}
\label{sec:imaging}

Following \cite{eckert14} we extracted count maps in the energy range $0.7-1.2$ keV (where the signal-to-background ratio is the highest) for the three EPIC detectors and co-added them to obtain a total EPIC image. To correct the images for vignetting, we extracted individual exposure maps using the XMMSAS task {\it eexpmap} and created a total exposure map by summing the individual exposure maps, weighted by the relative effective area of each instrument. Following \cite{snowden08}, we used a collection of closed-filter observations to produce a map of the non X-ray background, which we rescaled to our observations by comparing the count rates in the unexposed corners of the field of view. Finally, we created a vignetting-corrected, NXB-subtracted sky image by subtracting the NXB image from the count map and dividing by the exposure map. The resulting image was then smoothed using the XMMSAS adaptive-smoothing tool {\it asmooth}, ensuring a target signal-to-noise of 5. The final image of Hydra A is shown in Fig. \ref{fig:mosaic_xmm}, the newly discovered group is peaked on the galaxy LEDA 87445 and is shown in the figure by the white square.

\subsubsection{Spectral analysis}
\label{sec:spec}

\begin{figure}
\centerline{\resizebox{\hsize}{!}{\includegraphics[angle=270]{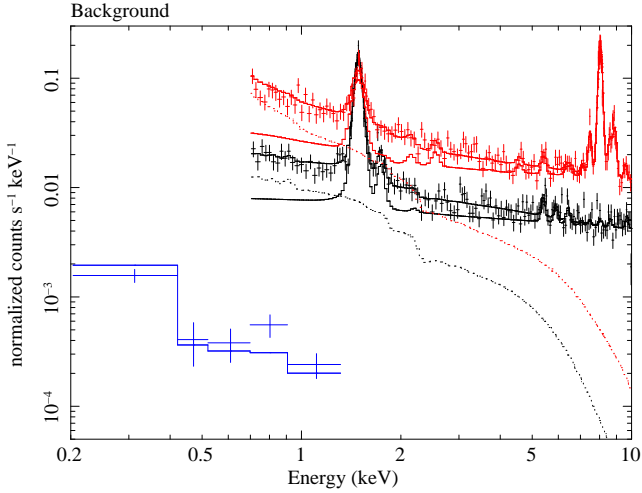}}}
\caption{Spectrum of the background region located at 25.5 arcmin SE of the cluster core used to measure the sky background components, from the EPIC pn (red), MOS2 (black) (MOS1 was omitted for clarity), and from the ROSAT/PSPC (blue). The solid lines show the NXB model for each instrument, while the dotted lines show the sky background model.}
\label{fig:bkg1_xmm}
\end{figure}

We performed a spectral analysis of the galaxy group using the the two good observations of the SE region in Hydra A
(Obs. ID. 0725240101 and 0761550101).
Spectra and response files for each region were extracted using the ESAS tasks {\it mos-spectra} and {\it pn-spectra} from the two observations separately, then the spectra of each detector were summed up using the {\it mathpha}, {\it addarf} and {\it addrmf} ftools v.6.16, and finally the total spectra were fit in XSPEC v12.8.1. 
Since the surface brightness of the group,  apart from the emission peak, barely exceeds the background level we preferred to model the background instead of subtracting it \citep[e.g.][]{leccardi08_t}.
This method requires a careful characterization of all the various background components to obtain reliable measurements of the relevant parameters.  We adopted the following approach to model the different spectral components:

\begin{itemize}
\item \textit{The non X-ray background (NXB):} we used closed-filter observations to estimate the spectrum of the NXB component in each region (ESAS tasks {\it mos-back} and {\it pn-back}), following the procedure described in \citet{snowden08}.
We approximated the NXB component by a phenomenological model, which we then included as an additive component in the spectral modelling \citep{leccardi08_t}.
The NXB components of the two MOS detectors are well reproduced by the models in \citet{leccardi08_t}, whereas the NXB spectrum of  the pn is best fitted by a double broken power-law model with photon indices fixed to 0.61 (below 1.68 keV), 0.31 (between 1.68 and 4.09 keV) and 0.18 (above 4.09 keV).  We left the normalisation  of the NXB component free to vary during the fitting procedure, which allows for possible systematic variations of the NXB level. The normalisation  of the prominent fluorescence  emission lines was also left free. The MOS observations are rather contaminated by quiescent soft protons (QSP) that survived the filtering process \citep[mean IN over OUT ratio of 1.32;][]{deluca04}, hence we included a broken power-law component with break energy at 5.0 keV, and slopes fixed to 0.4 (below 5 keV) and 0.8 (above 5 keV) \citep{leccardi08_t}, with the normalisation  left free to vary. We did not include this component in the analysis of the pn spectra, since the pn observation is only weakly contaminated by residual soft protons (IN over OUT ratio of 1.08).

\item \textit{The sky background components:} we used an offset region located $\sim 25.5$ arcmin away from the cluster core (see circle in Fig. \ref{fig:mosaic_xmm}), where no cluster emission is detected, to measure the sky background components in the region of Hydra A close to the accreting group.
To aid in constraining these components we added to the EPIC data the \emph{ROSAT} all-sky survey (RASS) background spectrum, sensitive to the 0.1-2.4 keV X-ray regime, and fitted the \emph{XMM-Newton} and RASS data jointly \citep{snowden97_backtool} . The RASS spectrum was obtained through the HEASARC X-ray background tool in a circular region of 0.15 degree radius overlapping the offset region used for the EPIC data.
The sky background was modelled  with three-components: \emph{i)} a power law with photon index fixed to 1.46 to model the cosmic X-ray background \citep[CXB,][]{deluca04}; \emph{ii)} a thermal component at a temperature of 0.22 keV to account for the Galactic halo emission; \emph{iii)} an unabsorbed thermal component at 0.11 keV for the local hot bubble. We fixed the CXB normalisation  to the value taken from \cite{deluca04} rescaled appropriately whereas we left the normalisation s of the other soft components free to vary. The normalisation s for the three background components are $(1.18\pm 0.20)\times10^{-6}$ arcmin$^{-2}$ (Local Bubble), $(7.15\pm 5.47)\times10^{-7}$ arcmin$^{-2}$ (Galactic Halo) and $9.81\times10^{-7}$ arcmin$^{-2}$ (CXB). For the APEC (Local Bubble and Galactic halo), these normalisation s are expressed in units of $10^{-14}/(4\pi d_A^2(1+z)^2)\int n_en_H\,dV$; for the power law (CXB), this is in units of photons keV$^{-1}$ cm$^{-2}$ s$^{-1}$ at 1 keV.
The best-fit spectrum for the offset region is shown in Fig. \ref{fig:bkg1_xmm}.
As a check we left free to vary in our model also the CXB normalisation . We found that its best-fit value, $(9.46\pm 1.08)\times10^{-7}$ arcmin$^{-2}$, agrees well within the statistical uncertainties with the expected value from \cite{deluca04}.
To model the sky background in a different region, the normalisation  of each component was rescaled by the ratio of the corresponding areas, accounting for CCD gaps and bad pixels.

\item \textit{The source:} we modelled the diffuse source emission in each region using the thin-plasma emission code APEC \citep{smith01_apec}, leaving temperature, metal abundance and normalisation  as free parameters \citep[the solar abundances were taken from][]{anders89} and the redshift was fixed to the optical value of the LEDA 87445 galaxy (z=0.0548, see Sect. \ref{sec:dynamics}). This component is absorbed by the Galactic hydrogen column density along the line of sight, which we fix to the 21cm value \citep[$N_H=4.54\times10^{20}$ cm$^{-2}$,][]{kaberla05}.
\end{itemize}

The spectral energy range considered is $0.7-10.0$ keV. We always used in the spectral fitting procedure the C-statistics rather than the $\chi^2$ \citep{cash79}, which is preferable for low S/N spectra \citep*{nousek89}. To apply the Cash statistic we performed a minimal grouping of the spectra to avoid channels with no counts.

\begin{figure}
\centerline{\resizebox{\hsize}{!}{\includegraphics[]{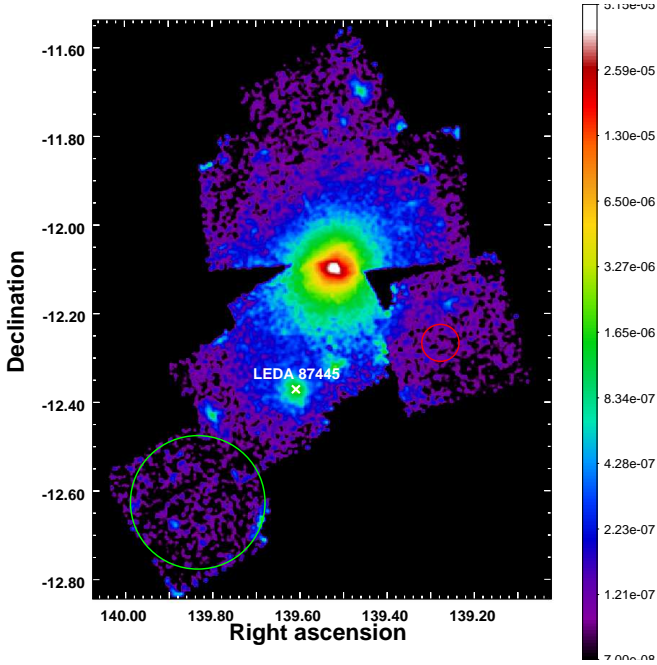}}}
\caption{Exposure-corrected \emph{Suzaku}/XIS mosaic image of Hydra A combining all seven available pointings. The LEDA 87445 group is highlighted in white. The green circle indicates the background region on the off-source pointing used for this analysis. The red circle is a region used to measure the ICM temperature of Hydra A, details are given in Sec. \ref{sec:specreg}.}
\label{fig:suzaku_mosaic}
\end{figure}

\subsection{\emph{Suzaku}}
\label{sec:suzaku}

\subsubsection{Data reduction}
\label{sec:dataredsuz}

Hydra A was observed by \emph{Suzaku} in 2012 for a total of 160 ks (PI: Eckert, ID 807087-807091) in addition to two pre-existing \emph{Suzaku} pointings \citep[70 ks, ID  805007-8; see][]{sato12}. This corresponds to a total of 6 pointings covering the entire azimuth of the cluster out to $R_{200}$, plus an off-source pointing located $\sim$ 40 arcmin from the cluster core. We processed all the observations using the \emph{Suzaku} FTOOLS as provided in HEAsoft v6.12 and the \emph{Suzaku} CALDB v20130305. We reprocessed the event files using the tool {\it aepipeline}, restricting to observing periods with cut-off rigidity COR$>$6. \emph{Suzaku}/XIS events obtained in 3x3 and 5x5 editing modes were combined using {\it Xselect}.

\subsubsection{Imaging analysis}
\label{sec:imagingsuz}

For each observation, we extracted photon images in the [0.5-2] keV band for the two front-illuminated chips (XIS0 and XIS3) and for the back-illuminated CCD (XIS1) and created exposure maps using the {\it xisexpmapgen} tool. We then produced an exposure-corrected mosaic image by combining all three detectors and all observations together. The resulting mosaic image is shown in Fig. \ref{fig:suzaku_mosaic}. The infalling group around LEDA 87445 is highlighted on the mosaic, together with the position of the background pointing.

\begin{figure}
\centerline{\resizebox{\hsize}{!}{\includegraphics[angle=270]{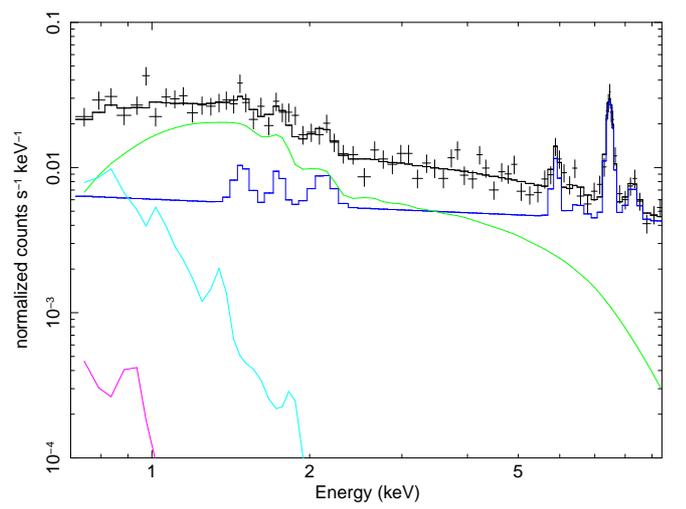}}}
\caption{\emph{Suzaku} front-illuminated spectrum of the background region shown in Fig. \ref{fig:suzaku_mosaic} and best-fit model (black line). The dark blue curve shows the NXB model, while the sky background components are shown in green (CXB), cyan (Galactic halo), and purple (local bubble).}
\label{fig:skybkg}
\end{figure}

\subsubsection{Spectral analysis}
\label{sec:specsuz}

We used {\it Xselect} to extract spectra in the regions of interest. We used the tools {\it xisrmfgen} and {\it xissimarfgen} to extract the appropriate response files for our regions, The data of the two front-illuminated CCDs were combined, while the XIS1 spectra were considered independently and fitted jointly. The non X-ray background spectra were estimated from the dark-Earth data using {\it xisnxbgen}. Instead of subtracting the NXB from the data, we fitted the NXB spectra using a phenomenological model and added this model as an additive component for the fitting procedure, similar to what explained in Sect. \ref{sec:spec} for \emph{XMM-Newton}.
To the best of our knowledge this is the first time the analysis of cluster outskirts is carried out by modelling  the \emph{Suzaku} NXB rather than subtracting it.
The sky background components were measured using the off-source pointing (see Fig. \ref{fig:suzaku_mosaic}) using the same model as described in Sect. \ref{sec:spec}.
The best-fit results for the sky components are $(1.11\pm0.13)\times10^{-6}$ arcmin$^{-2}$ (Local Bubble)
$(2.8_{-0.8}^{+2.9})\times10^{-7}$ arcmin$^{-2}$ (Galactic halo) with a Halo temperature of $0.33_{-0.09}^{+0.06}$ keV, and $(7.5\pm0.3)\times10^{-7}$ arcmin$^{-2}$ (CXB). For the APEC models (Local Bubble and Galactic halo), normalisation s correspond to $10^{-14}/(4\pi d_A^2(1+z)^2)\int n_en_H\,dV$; for the power law (CXB), the normalisation  is in units of photons keV$^{-1}$ cm$^{-2}$ s$^{-1}$ at 1 keV.

\begin{figure}
\centerline{\includegraphics[width=10.cm]{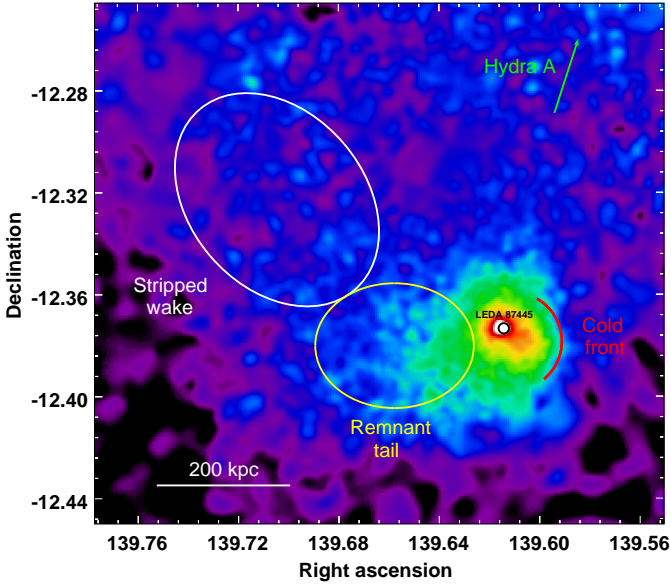}}
\caption{Zoom of the \emph{XMM-Newton} field on the LEDA 87445 group. Several noticeable features (central galaxy, cold front, remnant tail, stripped wake) are highlighted.} 
\label{fig:features}
\end{figure}

\begin{figure}
\centerline{\includegraphics[width=9.cm]{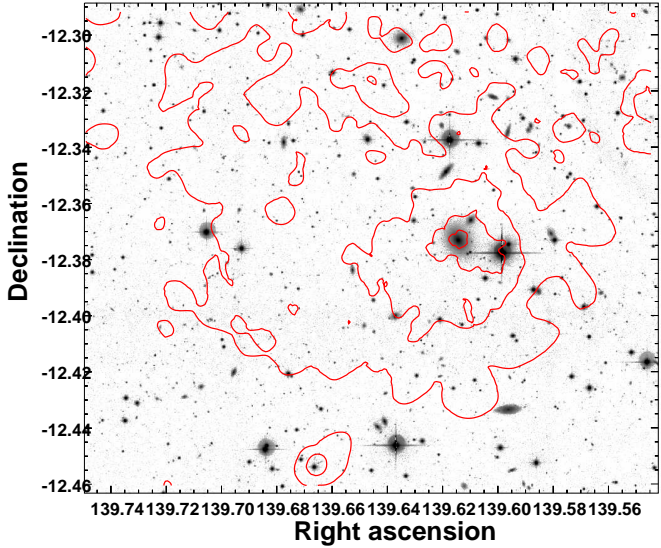}}
\caption{CFHT/Megacam r-band image of the region around the LEDA 87445 group with the X-ray contours overlaid in red.}
\label{fig:mega}
\end{figure}

\section{Results}
\label{sec:results}

\subsection{Morphology of the stripped group}
\label{sec:morph}

In  Fig. \ref{fig:features} we show a zoom of the \emph{XMM-Newton} X-ray emission around the accreting group situated 17 arcmin (1.1 Mpc) south-east from the Hydra A cluster center, and in Fig. \ref{fig:mega} we show a CFHT/Megacam r-band image of the same region with the X-ray contours overlaid.
A wide variety of features are revealed.
Coincident with the X-ray emission peak  there is the galaxy LEDA 87445 with redshift 0.05745 \citep[][]{smith04} similar to the Hydra A cluster redshift. East/north-east of the group emission peak we clearly detected a long excess of diffuse emission with projected length of  $\sim 12$ arcmin ($760$ kpc); this huge tail bends progressively from east towards the north-east.

\begin{figure}
\centerline{\includegraphics[width=9.cm]{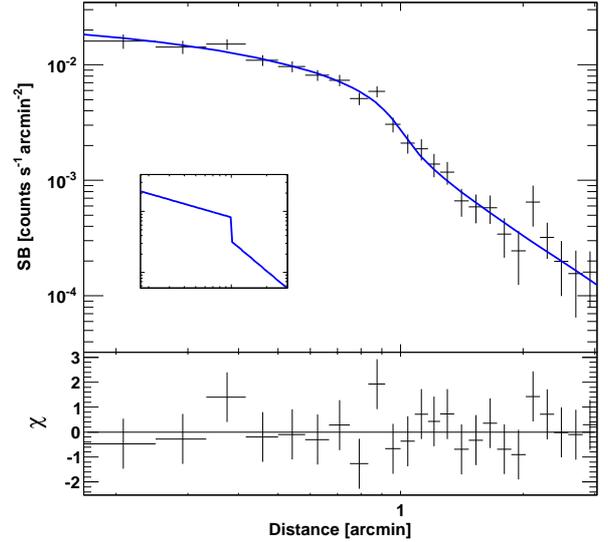}}
\caption{EPIC surface-brightness profile from the tip of the LEDA 87445 group in a narrow westward sector corresponding with the cold front. The solid line is the best-fitting broken power law (EPIC-PSF modelling  is included). The inset shows the corresponding deprojected density model.}
\label{fig:sb}
\end{figure}

\begin{figure*}
\resizebox{\hsize}{!}{\vbox{\hbox{
\includegraphics[angle=0,width=10cm]{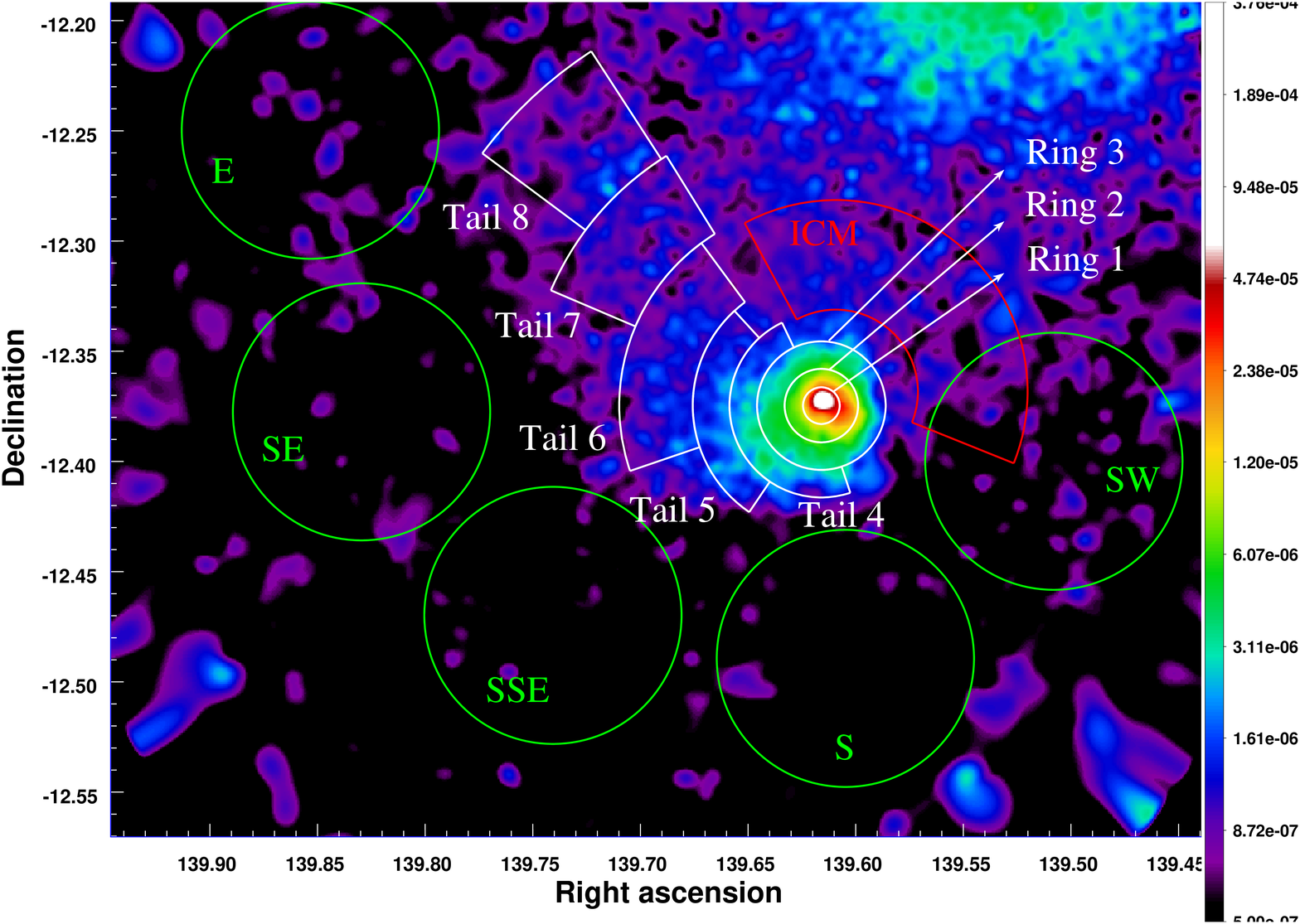}
\includegraphics[angle=0,width=9cm]{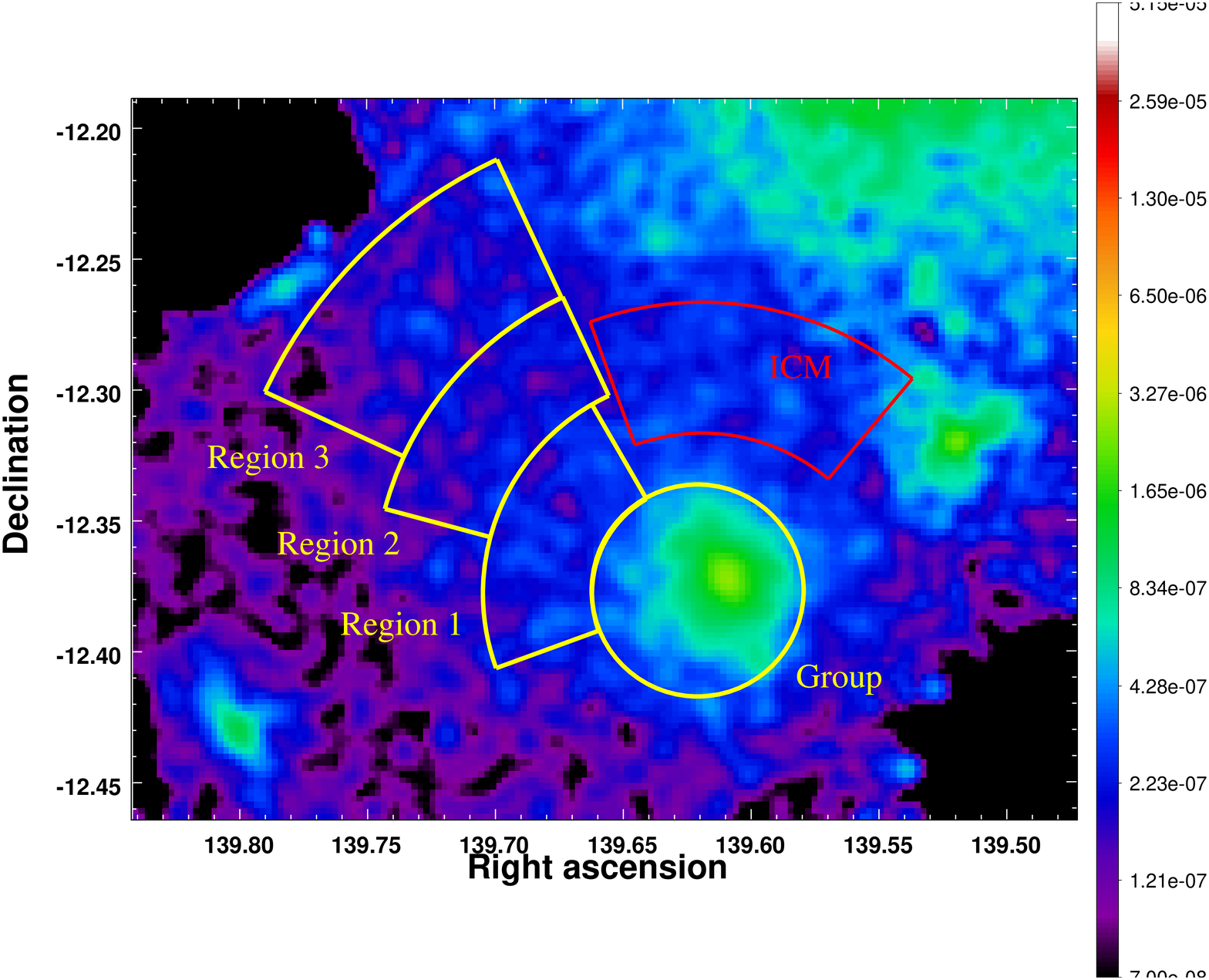}}}}
\caption{{\it Left panel:} {\it XMM-Newton} image of the LEDA 87445 group with the extraction regions used for the spectral analysis. White regions are chosen to match the LEDA group and its long tail of faint emission, the red sector is used to measure the ICM temperature nearby the group and the green circles are used to estimate the surface brightness of the ICM at various positions around LEDA. {\it Right panel:}  {\it Suzaku} image of the LEDA group with in yellow and red the regions used for the spectral analysis of the structure and of the ICM respectively. In both panels the labels of the regions are the same used in the Tab. \ref{tab:1} and \ref{tab:2} reporting the spectral analysis results. }
\label{fig:regions_all}
\end{figure*}

A sharp surface-brightness drop can be observed $\sim 1.5$ arcmin (100 kpc) west of the central galaxy. To confirm this statement, we extracted the surface brightness profile from the X-ray peak of the group in the west direction in a sector with position angles between $-60^\circ$  and $+30^\circ$ (measured counter--clock--wise from west) using the method described in \cite{eckert11} (see Fig. \ref{fig:sb}). A clear break in the surface-brightness profile can be seen 1.5 arcmin from LEDA 87445, which indicates the presence of a contact density discontinuity. We modelled  the surface-brightness profile using a broken power law density profile projected along the line of sight \citep{owers09} and convolved with the \emph{XMM-Newon} PSF \citep[see ][]{rossetti13}.
This model provides a good fit to the data ($\chi^2=14.25/19$ d.o.f.) and returns a density jump $n_{\rm in}/n_{\rm out}=2.5_{-0.6}^{+0.8}$.

\begin{figure}
\centerline{\resizebox{\hsize}{!}{\includegraphics[angle=90]{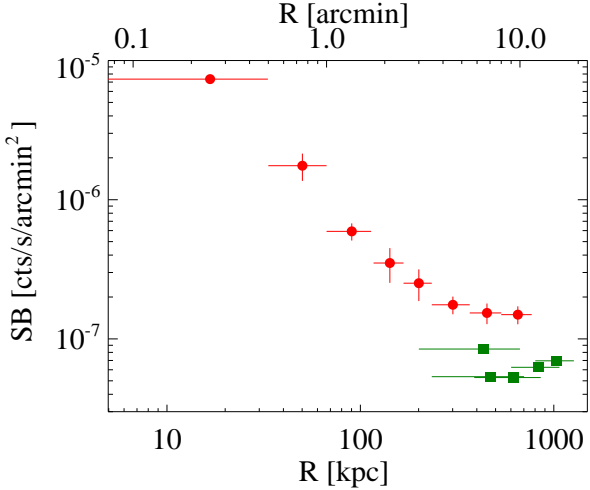}}}
\caption{EPIC surface-brightness in the [0.7-1.2] keV band for the group and tail regions (red circles that correspond to the white regions in Fig. \ref{fig:regions_all}) and for five regions surrounding the group (green squares that correspond to the green regions in Fig. \ref{fig:regions_all}).
}
\label{fig:sb_tail}
\end{figure}

\subsection{Spectral properties}
\label{sec:specreg}

We extracted EPIC spectra of the group and its long tail from the regions shown in Fig. \ref{fig:regions_all} (left panel) and analysed them following the prescription given in Sec. \ref{sec:data}. Spectra in similar regions were extracted also from the \emph{Suzaku} data (Fig. \ref{fig:regions_all}, right panel). The regions were chosen to trace the morphology of the observed feature as closely as possible.

\begin{figure*}
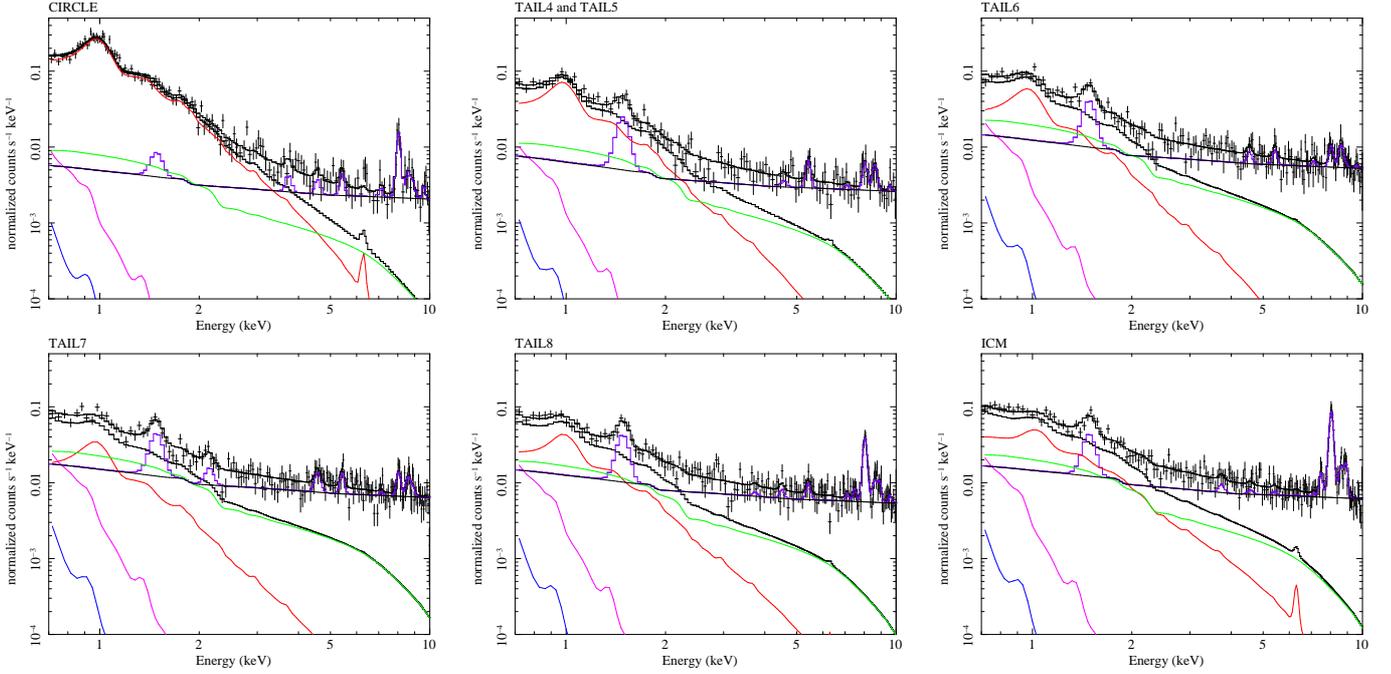

\resizebox{\hsize}{!}{\vbox{\hbox{
\includegraphics[angle=270]{fig10_circle_pn.eps}
\includegraphics[angle=270]{fig10_tail45_pn.eps}
\includegraphics[angle=270]{fig10_tail6_pn.eps}}
\hbox{
\includegraphics[angle=270]{fig10_tail7_pn.eps}
\includegraphics[angle=270]{fig10_tail8_pn.eps}
\includegraphics[angle=270]{fig10_icmhyd_pn.eps}}}}
\caption{EPIC/pn spectra for the regions defined in Fig. \ref{fig:regions_all} (the EPIC spectra were fitted simultaneously but here we show only the pn spectra and their spectral components for clarity proposes). The solid lines show the various components used for the fitting procedure: the source (red), the NXB (blue), the CXB (green), the Galactic halo emission (magenta), and the local hot bubble (cyan). \emph{Bottom right:} EPIC/pn spectrum of the local ICM in the region close to the tip of the infalling group.}
\label{fig:allspectra}
\end{figure*}

\begin{figure*}
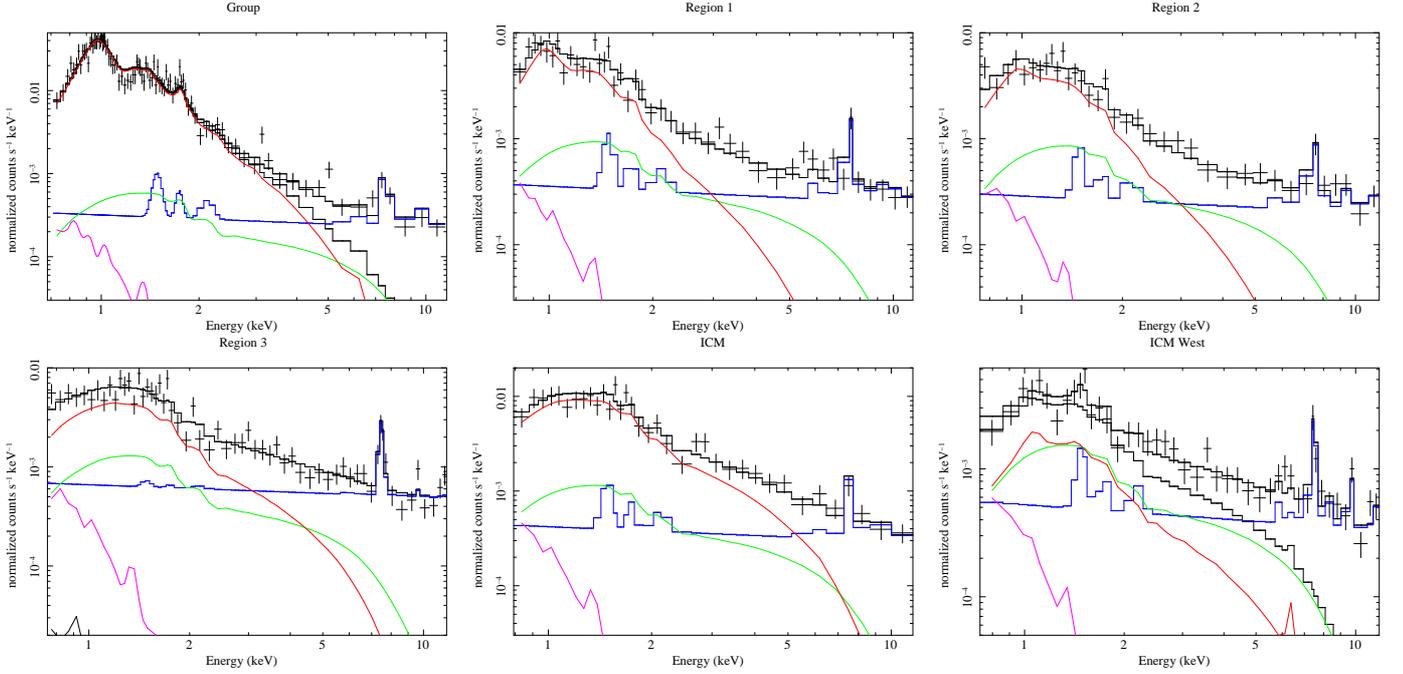

\resizebox{\hsize}{!}{\vbox{\hbox{
\includegraphics[angle=270]{fig11_group.ps}
\includegraphics[angle=270]{fig11_reg1.ps}
\includegraphics[angle=270]{fig11_reg2.ps}}
\hbox{
\includegraphics[angle=270]{fig11_reg3.ps}
\includegraphics[angle=270]{fig11_icm.ps}
\includegraphics[angle=270]{fig11_icm_west.ps}}}}
\caption{{\it Suzaku}/XIS spectra of the subgroup. The fit was performed using all 3 XIS detectors, however for clarity purposes we plot here only the spectra of the sum of the two FI units (XIS0+XIS3) and their spectral components. The solid lines show the various components used for the fitting procedure, the color code for the different spectral components is the same as in Fig. \ref{fig:allspectra}. The last two panels are the spectrum of the local ICM: the one labeled ICM refers to the region close to the tip of the infalling group and shown in Fig. \ref {fig:regions_all} (right panel), the other labeled ICM-SW is the measure in the circular region shown in red in Fig. \ref{fig:suzaku_mosaic}.
}
\label{fig:allspsuz}
\end{figure*}

The model fitting for each region was performed jointly on the 3 EPIC detectors for the \emph{XMM-Newton} data and on the 3 XIS detectors for the {\it Suzaku} data. In Fig. \ref{fig:allspectra} we display only the EPIC-pn spectra for clarity, whereas the XIS0 and XIS3 spectra were summed and we show the summed front-illuminated spectra in Fig. \ref{fig:allspsuz}. The best-fit parameters for the various regions are reported in Table \ref{tab:1} and Table \ref{tab:2}.

In the case of EPIC, we verified that the addition of an APEC spectral model in the {\it source} component (Sect \ref{sec:spec}), that takes into account the residual emission of the cluster Hydra A at the distance of the group, did not significantly change any result reported.
We also checked the robustness of our results against a possible cosmic variation of the CXB component in the EPIC spectra by varying the CXB normalisation by $\pm 50\%$, in the spectral models of each region. We found that the spectral parameters remain almost unchanged and always within the statistical $1-2\sigma$ uncertainties.

\begin{table*}
{\caption{\label{tab:1} XMM:  Best-fit parameters for the various regions defined in Fig. \ref{fig:regions_all} and for the local ICM around the tip of the group emission. See Fig. \ref{fig:allspectra} for the best fit and the various spectral components.}
\begin{center}
\begin{tabular}{lcccc}
\hline
Region & Distance [arcmin] & $kT$ [keV] & $Z_{\rm Fe}$ & $K$ [arcmin$^{-2}$]  \\
\hline
\hline
Circle    & $0.0   - 1.7$  & $1.33^{+0.01}_{-0.01} $ & $0.24^{+0.02}_{-0.01} $  & $6.4^{+0.2}_{-0.2}\times 10^{-5}$ \\
Ring 1   & $0.0   - 0.5$  & $1.27^{+0.01}_{-0.01} $ & $0.33^{+0.03}_{-0.03} $  & $2.4^{+0.1}_{-0.1}\times 10^{-4}$ \\
Ring 2   & $0.5   - 1.0$  & $1.36^{+0.01}_{-0.06} $ & $0.24^{+0.02}_{-0.02} $  & $1.0^{+0.1}_{-0.1}\times 10^{-4}$ \\
Ring 3   & $1.0   - 1.75$& $1.42^{+0.05}_{-0.05} $ & $0.19^{+0.03}_{-0.03} $  & $3.6^{+0.2}_{-0.2}\times 10^{-5}$ \\
Tail 4     & $1.75 - 2.5$  & $1.35^{+0.05}_{-0.03} $ & $0.18^{+0.03}_{-0.03}$  & $2.2^{+0.2}_{-0.2}\times 10^{-5}$ \\
Tail 5     & $2.5   - 3.5$  & $1.28^{+0.04}_{-0.05} $ &           $^{(a)}$                & $7.0^{+1.2}_{-1.1}\times 10^{-6}$ \\
Tail 6     & $3.5   - 5.5$  & $1.27^{+0.04}_{-0.05} $ & $0.14^{+0.05}_{-0.04}$  & $7.1^{+1.1}_{-1.1}\times 10^{-6}$ \\
Tail 7     & $5.5   - 8.0$  & $1.28^{+0.06}_{-0.08} $ & $0.09^{+0.04}_{-0.03}$  & $6.6^{+1.0}_{-0.9}\times 10^{-6}$ \\
Tail 8     & $8.0   - 11.5$& $1.25^{+0.09}_{-0.11} $ & $0.05^{+0.03}_{-0.02}$  & $1.3^{+0.3}_{-0.2}\times 10^{-5}$ \\
ICM$_{\rm Epic}$     & $2.2-5.2$ & $2.44^{+0.21}_{-0.20} $  &   $0.2$             & $8.8^{+0.6}_{-0.3}\times 10^{-6}$ \\
\hline
\end{tabular}
\end{center}
\textbf{Column description:} 1: Region as defined in Fig. \ref{fig:regions_all}. ``Circle" is the central circular region with radius 1.7 arcmin enclosing the regions Ring 1, 2 and 3. 2: Distance from the group peak in arcmin (1 kpc at $z_{\rm LEDA}=0.0575$ corresponds to 0.898 arsec). 3: Best-fit temperature in keV. 4: metal abundance ($^{(a)}$ the abundance reported for region Tail 4 in the table is measured from the summed spectra of regions Tail 4 and Tail 5). 5: normalisation  of the APEC model per arcmin$^2$.  The ICM$_{\rm Epic}$ region is at a projected distance from the cluster center of $\sim 15$ arcmin.
}
\end{table*}

Before proceeding with a detailed spectral analysis of the faint emission in the tail we decided to perform a crude but robust measure of this emission by comparing it to the cluster emission immediately south of the tail. From the  vignetting-corrected EPIC mosaic image of Hydra A (Fig. \ref{fig:mosaic_xmm}) we computed the mean surface brightness (SB) in the regions of the group and tail (white regions in the Fig. \ref{fig:regions_all}), and, for comparison, also in five regions nearby the group that are shown as green circles in the same figure. The results are plotted in Fig. \ref{fig:sb_tail}, where red circles are the SB from the tip of the group and along the tail, and squares are the SB of the nearby regions.
There is a clear evidence of an SB excess in the tail of the group with respect to the nearby background regions: if we consider two regions, one in the group and one in the background that are almost at the same (projected) radial distance from the core of the HydraA, such as for instance the regions named in Fig. \ref{fig:regions_all} Tail 8 and the SW, we find that the SB in the tail is higher by a factor of $\sim 2-3$ with respect to one measured in the background region.
This SB excess in the tail can be attributed only to an extra emission, since contaminations from the Galactic Halo foreground or residual ICM emission will affect all the considered regions at the same level.

Having established the high significance of the tail emission we now turn to the results of the spectral analysis to better characterise it. The normalisations obtained from the analysis of the EPIC and XIS data are significantly different (see Tab. \ref{tab:1} and \ref{tab:2}), with those measured by XIS higher by a factor $\sim 70\%$ than those measured by EPIC. This effect is likely due to the presence of the nearby bright Hydra A core just outside the \emph{Suzaku} field of view that produces ``stray light", and that contaminate our \emph{Suzaku} observation \citep{mori05}. To estimate the stray light effect we considered the \emph{Suzaku} region named ``Sector 3", where we measure  a flux of $8\times 10^{-14}$ ergs cm$^{-2}$ s$^{-1}$  (0.5-2 keV band). Considering that the total flux of Hydra A within 15 arcmin radius is $2.6\times 10^{-11}$ ergs cm$^{-2}$ s$^{-1}$ and that the stray light at this radius is $\sim 0.1\%$, the contribution from the Hydra A core in ``Sector 3" is roughly 1/3 of the flux.
If we add on top of that also the contribution from the LEDA 87445 group itself, which is less bright, but closer to the ``Sector 3" region, this very likely resolves the discrepancy between \emph{Suzaku} and \emph{XMM-Newton} normalizations.
The stray light could also be the cause of the lower metallicity measured with \emph{Suzaku} in the tail, since the extra emission would raise the observed continuum resulting in an underestimation of the metallicity.

\begin{table*}
\caption{\label{tab:2} SUZAKU: Best-fit parameters for the various regions defined in Fig. \ref{fig:regions_all} and for the local ICM around the tip of the emission.}
\begin{center}
\begin{tabular}{lccccc}
\hline
Region & Distance [arcmin] &$kT$ [keV] & $Z_{\rm Fe}$  & $K$ [arcmin$^{-2}]$ \\ 
\hline
\hline
Group        & $0-2.5$    & $1.26_{-0.03}^{+0.03}$ & $0.20_{-0.02}^{+0.03}$  & $1.11\pm 0.07\times 10^{-4}$ \\
Region 1    & $2.5-5$    & $1.23_{-0.08}^{+0.07}$ & $0.08_{-0.03}^{+0.04}$  & $2.13\pm 0.31\times 10^{-5}$ \\
Region 2    & $5-7.5$    & $1.24_{-0.14}^{+0.12}$ &Ê$0.04_{-0.02}^{+0.03}$  & $2.09\pm 0.39\times 10^{-5}$ \\
Region 3    & $7.5-11$  & $1.49_{-0.32}^{+0.20}$ & $0.06_{-0.04}^{+0.07}$  & $1.30\pm 0.19\times 10^{-5}$ \\
ICM            & $4.5-7.5$ & $2.68_{-0.31}^{+0.51}$ &   0.2                      & $1.56\pm 0.29\times 10^{-5}$ \\
ICM-SW     & $            $ &    $3.0_{-0.9}^{+1.3}$    &   0.2                     & $3.5\pm 0.4\times 10^{-6}$ \\
 \hline
\end{tabular}
\end{center}
\textbf{Column description:} 1: Region as defined in Fig. \ref{fig:regions_all}.  2: Distance from the group tip in arcmin. 3:  Best-fit temperature in keV. 4: metal abundance. 5: Normalisation of the APEC model per arcmin$^2$. 6: Minimum c-statistic and number of degrees of freedom.
{The ICM region is at the projected distance of $\sim 13$ arcmin to the cluster center (see right panel of Fig. \ref{fig:regions_all}); the ICM-SW regions is at the projected distance of $\sim 17.5$ arcmin south-west to the cluster center
(see red circle in Fig. \ref{fig:suzaku_mosaic})}.
\end{table*}

\subsubsection{Temperature of the ICM around the LEDA group}

We estimated the ICM temperature of Hydra A close to the LEDA group with both \emph{XMM-Newton} and \emph{Suzaku} data. The extraction regions for these measurements are shown in red in Fig. \ref{fig:regions_all}, they are at $\sim 15$ arcmin from the  Hydra A core for EPIC and $\sim 13$ arcmin from the core for XIS. We furthermore measured the ICM temperature from \emph{Suzaku} data in a region west of the Hydra A core at the same distance from the center as the LEDA galaxy (i.e., 17 arcmin), but far from the group emission (see red region in Fig. \ref{fig:suzaku_mosaic}).
The results are given in Tab. \ref{tab:1} and \ref{tab:2}.
The ICM temperature at such large cluster radii is difficult to measure because of the low surface brightness and high background level in the EPIC spectrum, and because of the stray light in that of the XIS.
Nevertheless, we find that the three ICM measures agrees within the statistical errors and are $\sim 2-3$ keV.

By comparing the normalisation per arcmin$^2$ of the two XIS, we note that the normalisation of the SW region is lower that those measured in the region close to the LEDA group. This difference is due to both to the larger distance from the Hydra A core, and, to the smaller "stray light" contribution in the west region (where the contaminating group is far away).
Moreover, the ICM normalisation measured with EPIC is higher than the one of the west region measured with XIS because of the presence of a surface-brightness excess upstream of the group motion, whose origin will be discussed in Sect. \ref{sec:disc}.

\subsection{Optical data and kinematical analysis}
\label{sec:dynamics}

Optically, Hydra A is a medium size cluster having Abell richness class $R=0$ \citep{abell89}. Few redshifts are reported in the literature. We analyzed the largest available homogeneous sample, that of 42 galaxies having line-of-sight LOS velocity ($V=cz$) published by \cite{smith04}. A detailed description of our analysis is given in Appendix \ref{app:a1}, below we report only the main results.

\begin{figure}
\centering
\includegraphics[width= 8cm]{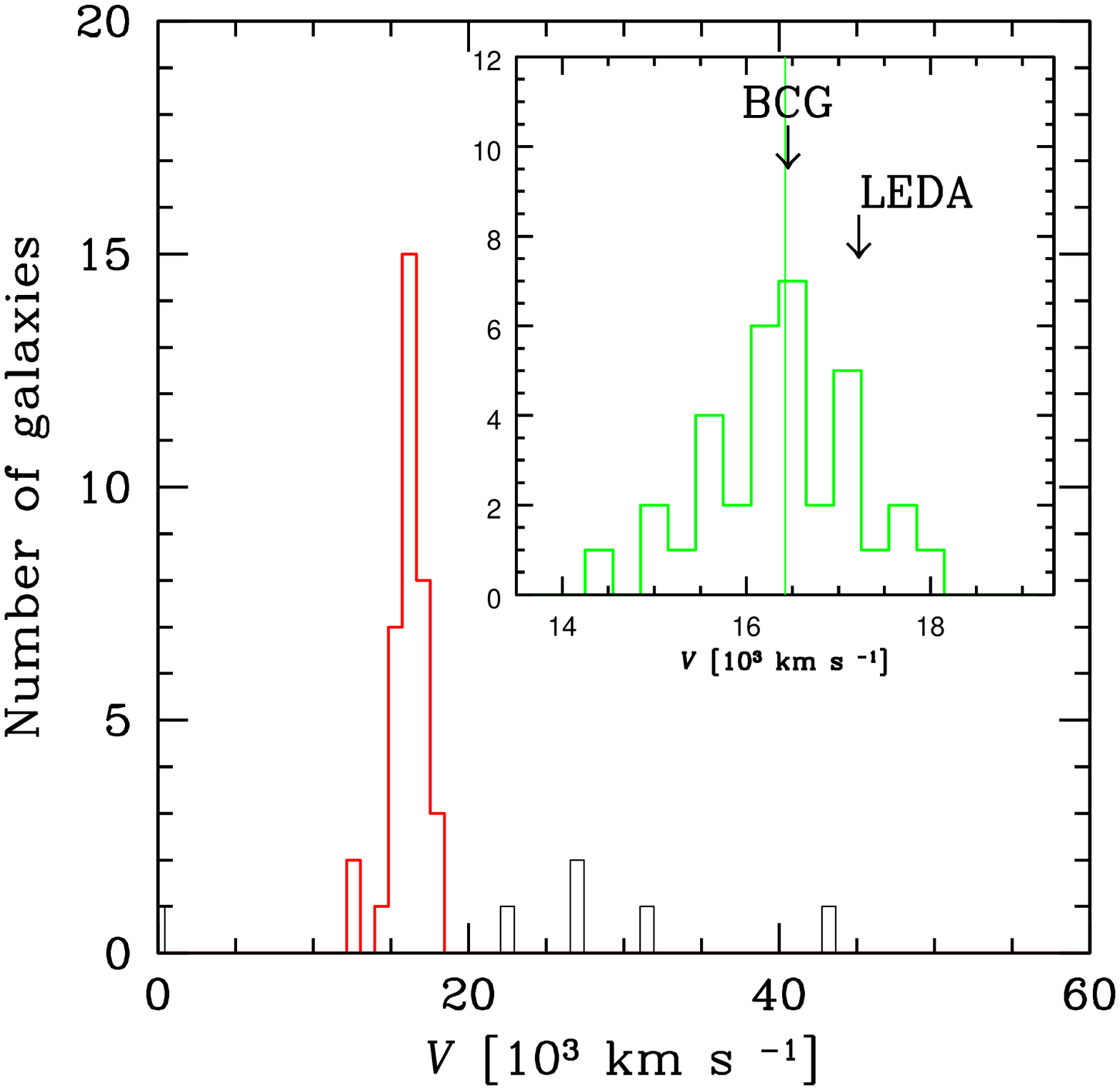}
\caption{The LOS velocity distribution of the sample of galaxies with spectroscopic redshift. The distribution of the 36 galaxies assigned to the redshift density peak of Hydra A is highlighted in heavy/red lines. The inset figure shows the distribution of the 33 final fiducial members. The mean cluster velocity (vertical faint green line) and the velocities of  Hydra A BGC and LEDA 87445 are indicated.  }
\label{fighisto}
\centering
\includegraphics[width=7.5cm]{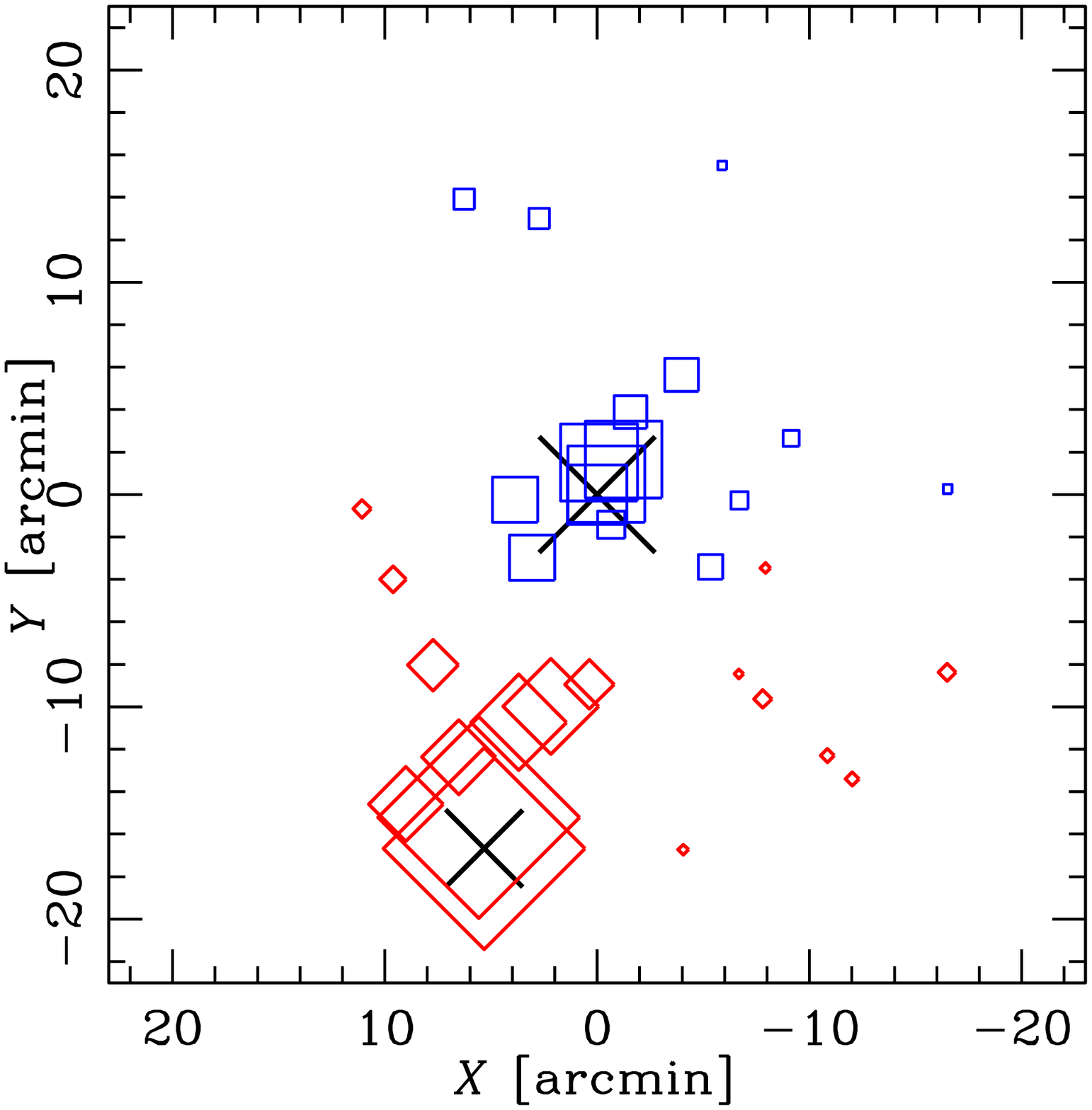}
\caption{DS$\left<V\right>$ bubble-plot for the 33 cluster members.  The larger the symbol, the larger is the deviation of the local mean velocity from the global mean velocity. Blue squares indicate negative deviations and red rotate squares positive deviations. Crosses indicate the BGC (center) and LEDA 87445.  }
\label{figdssegno5v}
\end{figure}

We have taken as the center of the Hydra A cluster the position of its brightest cluster galaxy (BCG), that has  a LOS velocity of $V_{\rm BCG}=16\,452$ \kss, and we have applied to the galaxies in the \cite{smith04} sample the DEDICA reconstruction method to investigate the LOS velocity distribution \citep[e.g.][and references therein]{girardi98}. In Fig.~\ref{fighisto} we plot the resulting sample of 36 galaxies assigned to the redshift density peak of Hydra A and the LOS velocity distribution of 33 fiducial cluster members, shown in the inset of the figure. The mean cluster redshift derived from the 33 fiducial members is $\left<z\right>=0.0548\pm0.0004$, i.e. $\left<V\right>=(16\,421\pm121)$ \ks and the LOS velocity dispersion is $\sigma_{V}=723_{-72}^{+93}$ \kss.

The bright galaxy LEDA 87445 is a cluster member lying in the high velocity part of the cluster-galaxies velocity distribution with a LOS velocity of $V_{\rm LEDA}=17\,224$ \kss. We applied a Dressler \& Schectman test \citep[DS-test]{dressler88}, to probe if the regions around the BCG and LEDA 87445 galaxies are populated by galaxies with different kinematics. This test allowed us to detect a substructure around LEDA 87445 with a c.l. larger than 99\%. In Fig.~\ref{figdssegno5v} we report the results for the DS-test based on the mean velocity (DS$\left<V\right>$) of the 33 cluster members. In the bubble-plot shown in the Fig. \ref{figdssegno5v}, the larger symbols represent larger deviations of the local mean velocity from the global mean velocity and the colors indicate negative (blue) and positive (red) deviations. The plot shows a clear concentration of galaxies with positive deviation around the LEDA 87445 galaxy. The existence of correlations between positions and velocities of cluster galaxies is always a strong footprint of real substructures.

We conclude that around the LEDA 87445 galaxy exists a group of galaxies and that this group is likely part of the Hydra A cluster halo.

\begin{figure*}
\resizebox{\hsize}{!}{\vbox{\hbox{
\includegraphics[angle=90,width=10cm]{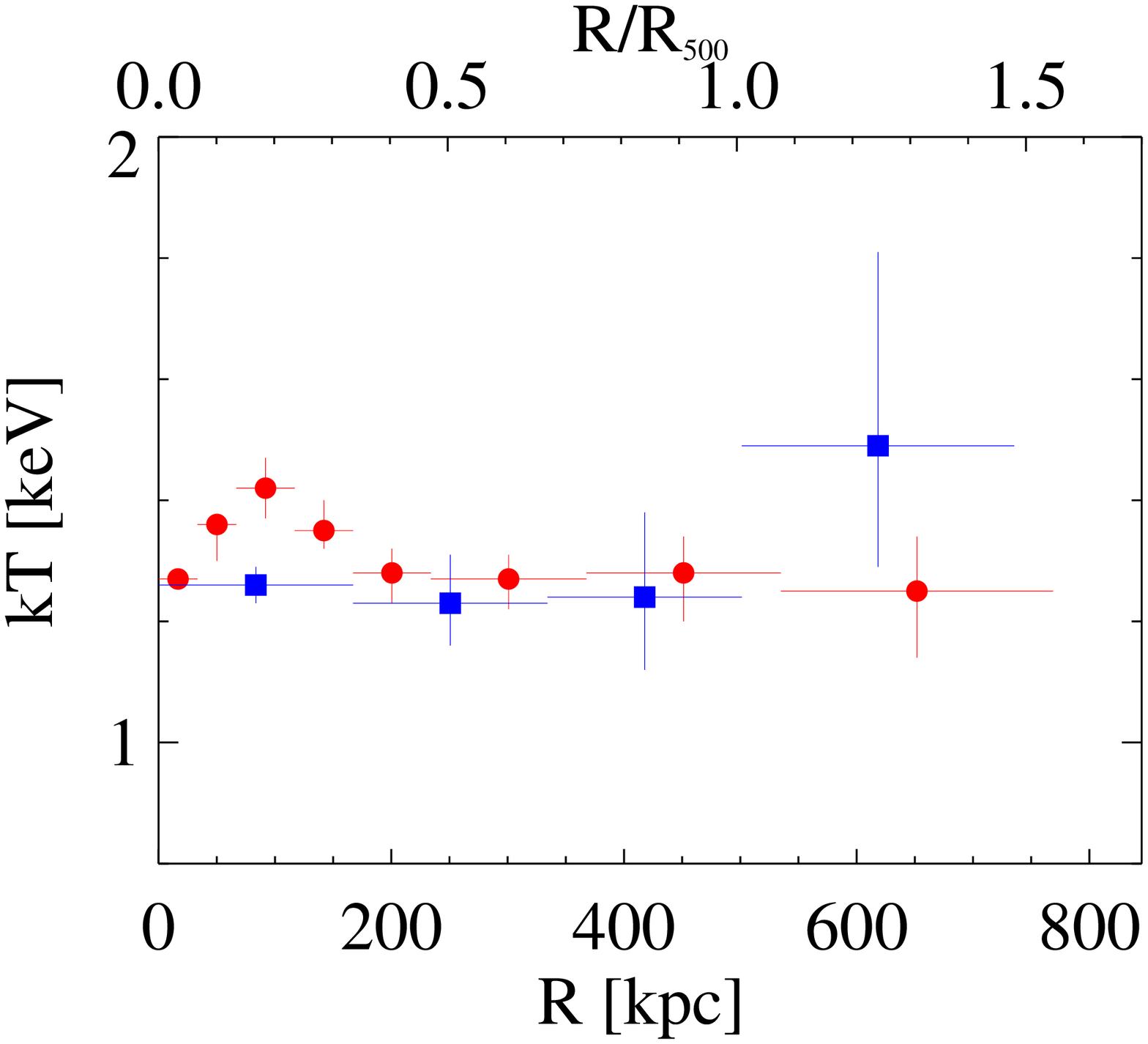}
\includegraphics[angle=90,width=10cm]{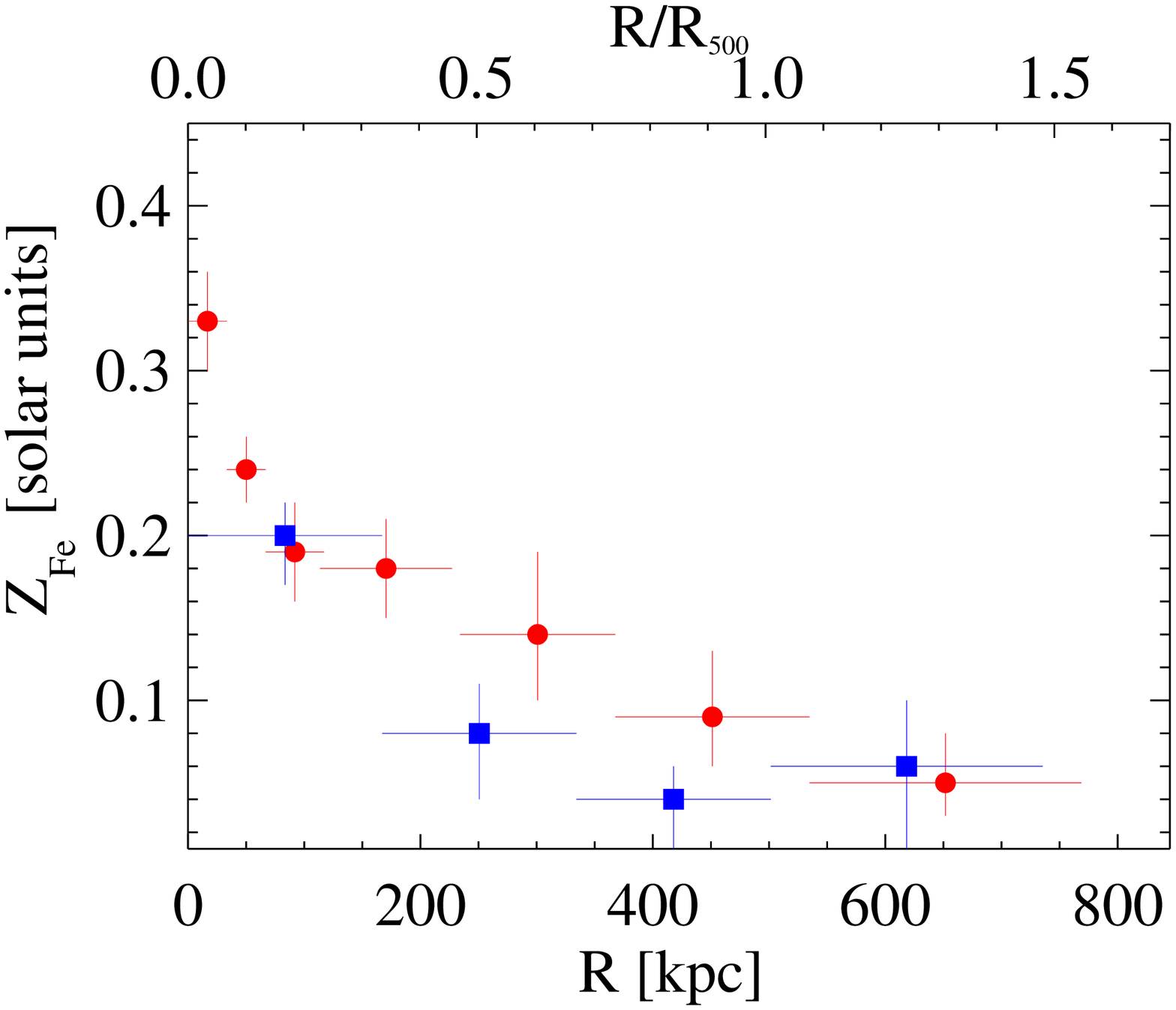}}}}
\caption{{\it Left panel}: Temperature profile of the LEDA 87445 group and tail. {\it Right panel}: Metal abundance profile.
In both panels filled red circles are EPIC data and filled blue squares are XIS data. The radius is the distance from the group emission peak of the regions shown in Fig. \ref {fig:regions_all}.}
\label{fig:profx}
\end{figure*}

\section{Discussion}
\label{sec:disc}

\subsection{The LEDA 87445 galaxy group}
\label{sec:int}

Our analysis of the correlation between the position and velocities of the galaxies available in the literature, presented in Sec.\ref{sec:dynamics}, has revealed that around LEDA 87445 exists a galaxy group with redshift similar to that of Hydra A. From the spectral analysis of the regions lying just inside and outside the sharp surface-brightness drop (Sect. \ref{sec:specreg}) we found temperatures kT$_{\rm in} = 1.33\pm0.01$ keV and kT$_{\rm out} = 2.44\pm0.20$ keV (see Table \ref{tab:1}) that allows us to conclude that this density discontinuity is a cold front produced by the motion of the galaxy group in the ambient ICM of Hydra A.

Within this picture the long diffuse emission north-east of the group peak can be explained in a natural way as a ram-pressure stripped tail of gas which initially belonged to the group. In Fig. \ref{fig:features} we can distinguish between the part of the tail directly connected with the emission peak and extending towards east (``remnant tail"), which is probably still gravitationally connected with the group peak emission, and the rest of the tail that points towards NE (``stripped wake''), probably formed by gas already stripped from the main group and ready to be mixed with the ICM. The common origin between the X-ray emitting gas in the tip and all along the tail is strengthened by the thermal properties of the gas in the whole region: the constant temperature (Fig. \ref{fig:profx}, left panel) up to $\sim 760$ kpc from the peak and its lower value with respect to the ICM temperature (Tab. \ref{tab:1} and \ref{tab:2}) indicates the same origin of the gas.

The core of the LEDA 87445 group appears well defined, indicating that we are observing the system likely during its first core passage, with a substantial part of the gas still gravitationally retained by the group.

\subsection{Thermodynamical properties of the LEDA group}
\label{sec:profx}

Temperature and metal abundance distributions in the group LEDA 87445 and its tail are plotted in Fig. \ref{fig:profx}, where red circles are EPIC and blue squares are XIS measurements.
Both profiles are centered on the X-ray emission peak of the structure (RA=9$^h$:18$^m$:27.783$^s$, DEC=-12$^o$:22$^\prime$:30.66$^{\prime\prime}$) and follow the faint tail as shown in Fig. \ref{fig:regions_all}. From the mean temperature of the group ($1.3$ keV) and the scaling relation of \cite{arnaud05} we estimated the R500 radius of the group, R500 = 497 kpc, and we used this value to rescale the profiles. EPIC and XIS measurements agrees rather well in both profiles. The slightly lower iron abundance found by XIS beyond 200 kpc is likely related to the stray-light and PSF effects of \emph{Suzaku}, already discussed in Sec. \ref{sec:specreg}.

We find that the temperature along the group is roughly constant at $\sim 1.3$ keV, which is a value typical of the virialized halo of galaxy groups with mass of a few $10^{13}$ M$_\odot$.
At the center kT is 1.27 keV then increase up to 1.42 keV at about 100 kpc and then decreases to 1.3 at $200-220$ kpc (which is the end of the remnant tail), beyond the temperature remain constant along all the stripped wake up to $>500$ kpc.

The metal abundance profile (Fig. \ref{fig:profx}, right panel) shows the typical trend observed in the core of relaxed groups and clusters, namely a higher abundance (0.33 Z$_{\rm Fe}$) at the emission peak which then decreases moving away from the peak \citep[e.g.][]{leccardi08_ab,degrandi04}. In the remnant tail, between $\sim 100-200$ kpc, the abundance reach the typical value for the outermost regions of group \citep[see review by ][i.e. $\sim 0.2$ Z$_{\rm Fe}$]{sun12_rev}, then declines even more along all the faint and patchy stripped wake to very low values.  The rather low abundance in the wake could be explained, as in the case of A2142 \citep{eckert14}, by a multi-temperature structure of the gas; in fact, when a multiphase gas with temperatures below $\sim 2$ keV is fit with a single temperature spectral model it returns an iron abundance biased towards low values. This is a well studied effect known as the Fe-bias \citep{buote00b,buote00a}.

The surface brightness fluctuations observed in the stripped gas of the faint and patchy tail (see Fig. \ref{fig:features}), likely related to gas clumpiness or to heated ICM gas mixed to the stripped gas, actually suggest that the gas is multiphase.

\begin{figure}
\centerline{\resizebox{\hsize}{!}{\includegraphics[angle=90]{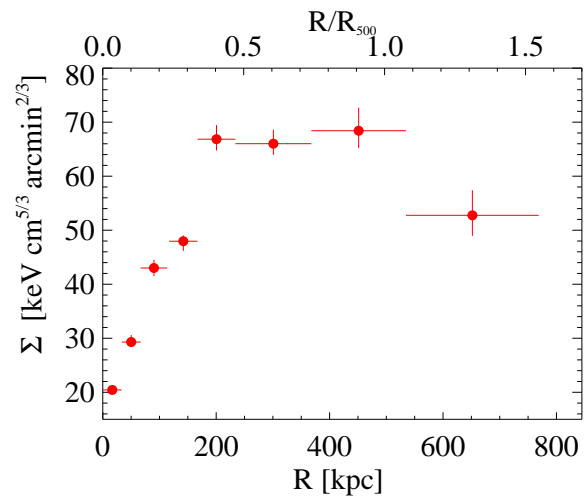}}}
\caption{Pseudo-entropy profile of the LEDA 87445 group and tail regions (EPIC data only).}
\label{fig:entprof}
\end{figure}

In Fig. \ref{fig:entprof} we show the pseudo-entropy profile of the gas, defined as $\Sigma = {\rm kT/K}^{1/3}$, where both kT and K are taken from Table \ref{tab:1}, along the tail of the group.
The profile is characterized by a rapid increase as we move away from the core, typical of cool core systems, followed by a first flat region between 60 kpc and  160 kpc, a  jump, around 160 kpc,  and a more extended flat region going from 160 to 500 kpc. The jump occurs  roughly where the tail starts to become clumpy and likely marks the transition between the remnant tail and the stripped tail. The shape of the profile suggests that the gas evolves adiabatically both in the region where it is gravitationally bound to the group and further out. However the transition between the two regimes is not adiabatic, there is a sudden and significant increase in entropy. We note that, since the gas appears to be more clumpy in the stripped than in the remnant tail, the density in the latter is likely more significantly overestimated than in the former, implying that the jump in entropy is even more pronounced than we estimate.  While the exact nature of the process responsible for heating the gas is unclear, its abrupt nature suggests that it is related to the stripping process itself. 

\subsection{ICM emission north-east  and north-west of the group}
\label{sec:north}

As shown in Figures  \ref{fig:features} and \ref{fig:regions_all}, the ICM emission north-west and north-east of the LEDA group is more intense than in other cluster regions located at the same radial distance from the center. If, as the position of the cold front suggests, the group is currently moving west, the upstream enhancement is likely due to compression by the group itself. Moreover if, as suggested by estimates provided in subsection \ref{sec:motion}, the motion is subsonic, the compression is adiabatic  and the entropy we measure in this region should be similar to that of neighbouring regions, where the lower surface brightness prevents direct measures.
Conversely, the enhancement north-east of the group is likely related to tidal effects. Indeed, as the group moves through the atmosphere of the cluster, gas will be displaced towards it \citep[see Fig.3, panels 3 to 5, in][]{ascasibar06}. 
It is worth noticing that the contiguity between the gas  in the tail (see Sect. \ref{sec:profx}) and that in the north-east excess makes it difficult to provide a  clear separation between the two. 

\subsection{Ram-pressure stripping properties}
\label{sec:ram}

The LEDA 87445 group has a peak and a rather long tail (760 kpc). We performed a  simple estimate of the velocity of the moving group by assuming that the gas at the tip of the structure is in pressure equilibrium with its surrounding ICM \citep{markevitch00}, \begin{equation} \label{eq:pram}  P_{\rm ICM } + \rho v^{2}_{\rm rel}  =  P_{\rm group}, \end{equation} where $v_{\rm rel}$ is the velocity with which the group is moving through the cluster, $\rho$ is the density of the ambient ICM and $P_{\rm ICM}$ and $P_{\rm group}$ are  the pressure associated to the ambient ICM and the tip of the group, respectively. We began by estimating gas densities on the two sides of the cold front  from the surface brightness profile (see Fig. \ref{fig:sb}). From the spectral analysis reported in Sect. \ref{sec:specreg}, we measured temperatures of 1.3 keV for the for the downstream region and 2.44 keV for the ICM region. By combining density and temperature measures we derived estimates for $P_{\rm ICM }$ and $P_{\rm group} $. Finally by applying Eq. \ref{eq:pram} we come up with an estimated velocity of $440\pm 170$ km/s, which about half the speed of sound, $c_s=\sqrt{\frac{\gamma kT}{\mu m_p}}=820\pm60$ km/s, for a plasma of  2.44 keV.

The detection of a cold front west of the group center allowed us to estimate the velocity of the moving group  through the stagnation point argument developed by \cite{vikhlinin01a}. This estimate should return a more precise estimate of the velocity, since it is using the measurement of the density jump directly at the front. From the measurement of the density jump at the front ($n_{\rm in}/n_{\rm out}=2.5_{-0.6}^{+0.8}$, see Sect. \ref{sec:morph}) and the temperature difference between the regions inside and outside the front (see Table \ref{tab:1}), we estimate a pressure jump $p_{\rm in}/p_{\rm out}=1.4_{-0.3}^{+0.5}$, which through the stagnation point argument corresponds to a Mach number $\mathcal{M}=0.6_{-0.4}^{+0.5}$ \citep[see][]{vikhlinin01a,markevitch07_rev}. Given the sound speed in the medium beyond the cold front, we estimate that the group is moving at a speed of $510_{-340}^{+430}$ km/s relative to the main cluster.

Thus, from both the estimates of the velocity, the group is falling onto the main cluster at, roughly,  half the sound speed . Under the assumption that the group has been moving at a constant velocity we find that the gas in the outermost region of the structure was expelled some $\sim 1.4-1.6$ Gyr ago.

Assuming that the extension of the emission on the line of sight is the same as in the plane of the sky, we derived a crude estimate of the gas mass that resides in the tail from the normalisations of  "Tail" regions 4 to 8 (see Tab. \ref{tab:1}), finding $1.4\times 10^{11}$ M$_{\odot}$. A similar analysis for the "Ring" regions 1 to 3 leads to a mass of $1.1\times 10^{11}$ M$_\odot$ for the tip of the structure. Thus our analysis seems to indicate that roughly 60\% of the  observed gas mass is located in the tail. From the temperature of the substructure we estimate a total mass of the order of a few $(2-5)\times 10^{13}$ M$_\odot$  \citep{sun09}.  Assuming a gas fraction of 5\% \citep{sun09} a group like ours should have a gas mass of the order of $1.5\times 10^{12}$ M$_\odot$, which is roughly a factor of 6 larger than what we estimate to be currently associated with either the group or its tail. Assuming that all the excess emission observed in the Hydra A SE region, is associated with gas once residing in the groups potential well we come to a conservative upper limit for the total gas mass of the group of $5\times 10^{11}$ M$_{\odot}$, which is still low by a factor of 2.5 with respect to expectations. While gas mass estimates for groups are clearly less precise than those for clusters, this results seems to suggest that our group may have already lost a significant fraction of gas prior to the stripping event we are currently observing.
Irrespective of what the starting gas mass of our group might have been, the current data is already sufficient to inform us that a significant fraction of the gas has been stripped from the group. Assuming that the part of the tail nearest the tip be made of gas that is still gravitationally bound to the group, we estimate the stripped gas to be  between 1 and 5$\times 10^{11}$ M$_\odot$, i.e. half and four fifths of the total gas mass, depending on whether the excess gas  in the Hydra A SE region was donated by the group.

As in the case of the group falling onto A2142 \citep{eckert14}, we have performed an estimate of the timescale required for the gas in the tail to be heated up to the temperature of the ambient ICM. We have assumed 100 kpc for the length scale over which the temperature gradient develops; we have also taken an electron density for the ICM surrounding the tail of  $7 \times 10^{-5}$ cm$^{-3}$ (this is estimated from the ROSAT surface brightness profile of Hydra A, provided by  \cite{eckert12}, by excluding the sector contaminated by the group emission) and an ICM temperature of 2.44 keV.
Finally, adopting Eq. 2 and 3 of \cite{eckert14} (originally derived in \citealt{gaspari13}), we have found a conduction timescale of 21.1$f^{-1}$ Myr, where $f$ is the effective isotropic conduction suppression factor. By equating this timescale with the age of the structure, that was previously determined to be between 1.4 and 1.6 Gyr, we estimate a suppression factor of more than 70. Thus, as for the case of A2142, we find evidence for significantly suppressed conductivity.

We note that the lack of significant heating of the gas in the tail is in agreement with the flat entropy profile we measure along the stripped tail (Fig. \ref{fig:entprof}). Indeed, if heating were taking place, we would expect an increase in entropy as we move away from the group.
The high suppression of conduction is consistent with the survival of cold fronts and filamentary structures \citep{ettori00,forman07,sanders13}. Through the power spectrum of density perturbations, \cite{gaspari13,gaspari14} also retrieved a high suppression factor in the Coma cluster. Another recent example showing the suppression of transport processes is given by \cite{ichinohe15} in their study of the complex structure of cluster Abell 85. The presence of quasi-linear high-density filamentary structures spanning several kpc observed in the cores of nearby clusters, such as the Coma cluster \citep{sanders13_sci}, consisting of low-entropy material that was probably stripped from merging subclusters, shows that the conduction is heavily suppressed also in cluster centers. Most recent observations seem thus to agree on the low thermal conductivity of the ICM. Magnetic fields are presumably the most effective agents to maintain these features separate from the ICM.

\begin{figure}
\centerline{\includegraphics[angle=270,width=9cm]{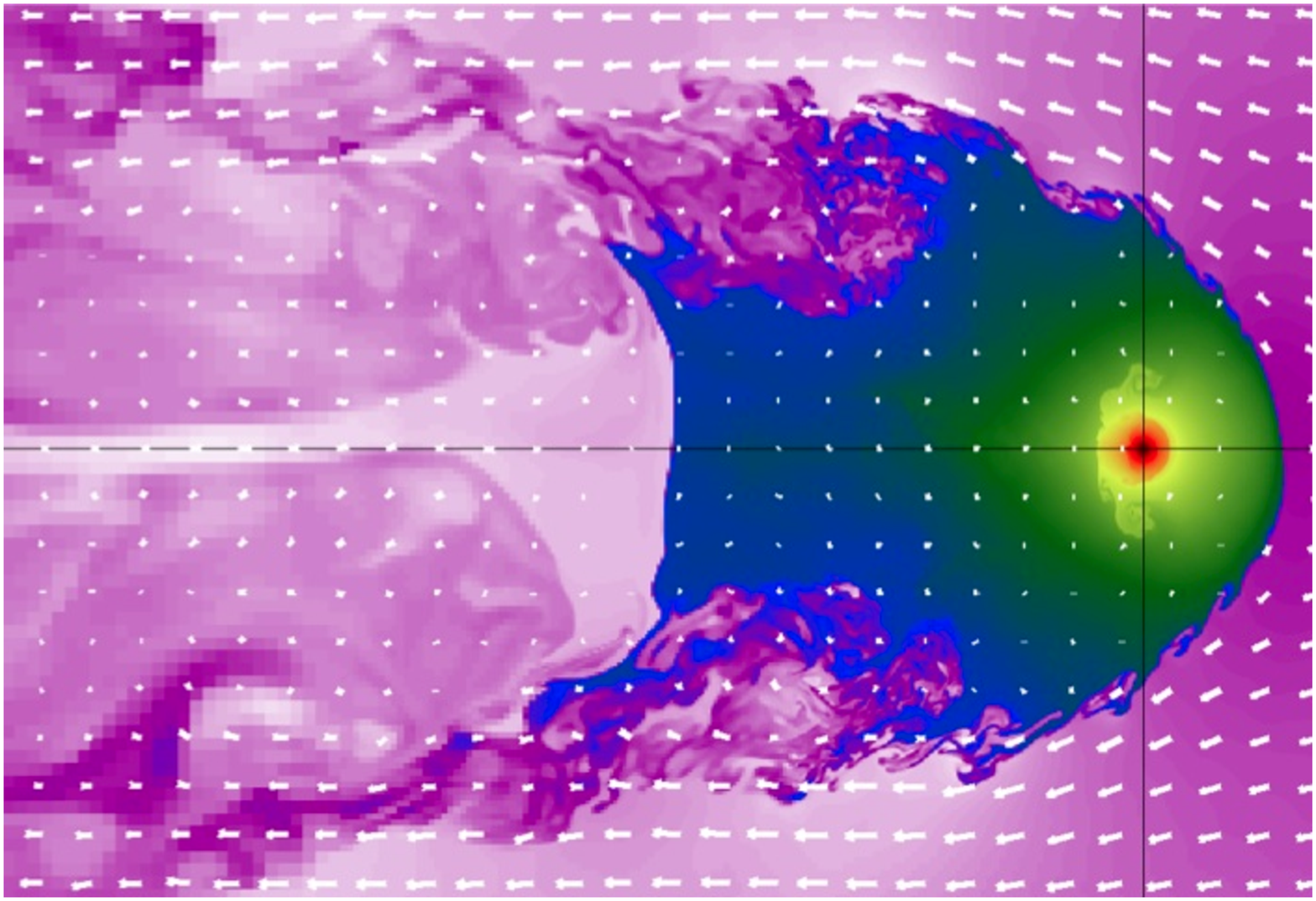}}
\caption{From \cite{roediger15a}. Snapshot of a high-resolution simulation of the hot atmosphere of an elliptical galaxy falling into a galaxy cluster. The bottom of the potential well is marked with the black cross. The arrows show the velocity field of the fluid. Several features can be seen: the cold front at the peak with the KH rolls on the side, the Òremnant tailÓ in the blue low-velocity area, and the stripped tail trailing behind.}
\label{fig:simul}
\end{figure}

\subsection{Infalling group motion}
\label{sec:motion}

In the previous section we have assumed that the motion of the accreting group is on the plane of the sky. This is suggested in the X-rays by the position of the cold front pointing towards west, whereas the main Hydra A cluster is located in the north direction, and by the shapes of tail and wake, that gradually curve from the east to the north (see Fig. \ref{fig:features}).
We postulated that the group was originally coming from the north-east direction with a large impact parameter and that its trajectory is being bent by the gravitational pull of Hydra A.

The stagnation point analysis lead to a velocity of the group of the order of half of the sound speed, this is the 3D velocity in the hypothesis that the orbit is not inclined too much out of the plane of the sky.
However, the LOS velocity difference between the LEDA 87445 galaxy and the BCG of Hydra A,  772 km/s, is of the same order of the cold front velocity, implying that the group is likely moving with a certain angle along the LOS, i.e $\sim 30^\circ-50^\circ$.
The visibility of the cold front would be preserved with such an orbital angle. \cite{roediger15b} show mock X-ray images of inclined galaxies falling into a group/cluster ICM (see their Figure 7 and 12) where the cold front is visible at least up to inclinations of 60 degree out of the plane of the sky. If the cold front is indeed inclined by $\sim 45^\circ$ with respect to the plane of the sky, then projection effects will likely lead to an underestimation of the group velocity from the stagnation point analysis.

To fully understand the motion of the group more detailed X-ray and optical observations, as well as detailed analysis on mock observations, would be highly useful.

\subsection{Comparison with numerical simulations}
\label{sec:simul}

Ram pressure stripped, X-ray emitting wakes and tails are observed trailing galaxies infalling onto groups and clusters. Several recent works have addressed the subject with numerical simulations \citep[e.g.][and references therein]{cen14,vijaya15,roediger15a,roediger15b}.

\begin{figure}
\centerline{{\includegraphics[angle=270,width=13.5truecm]{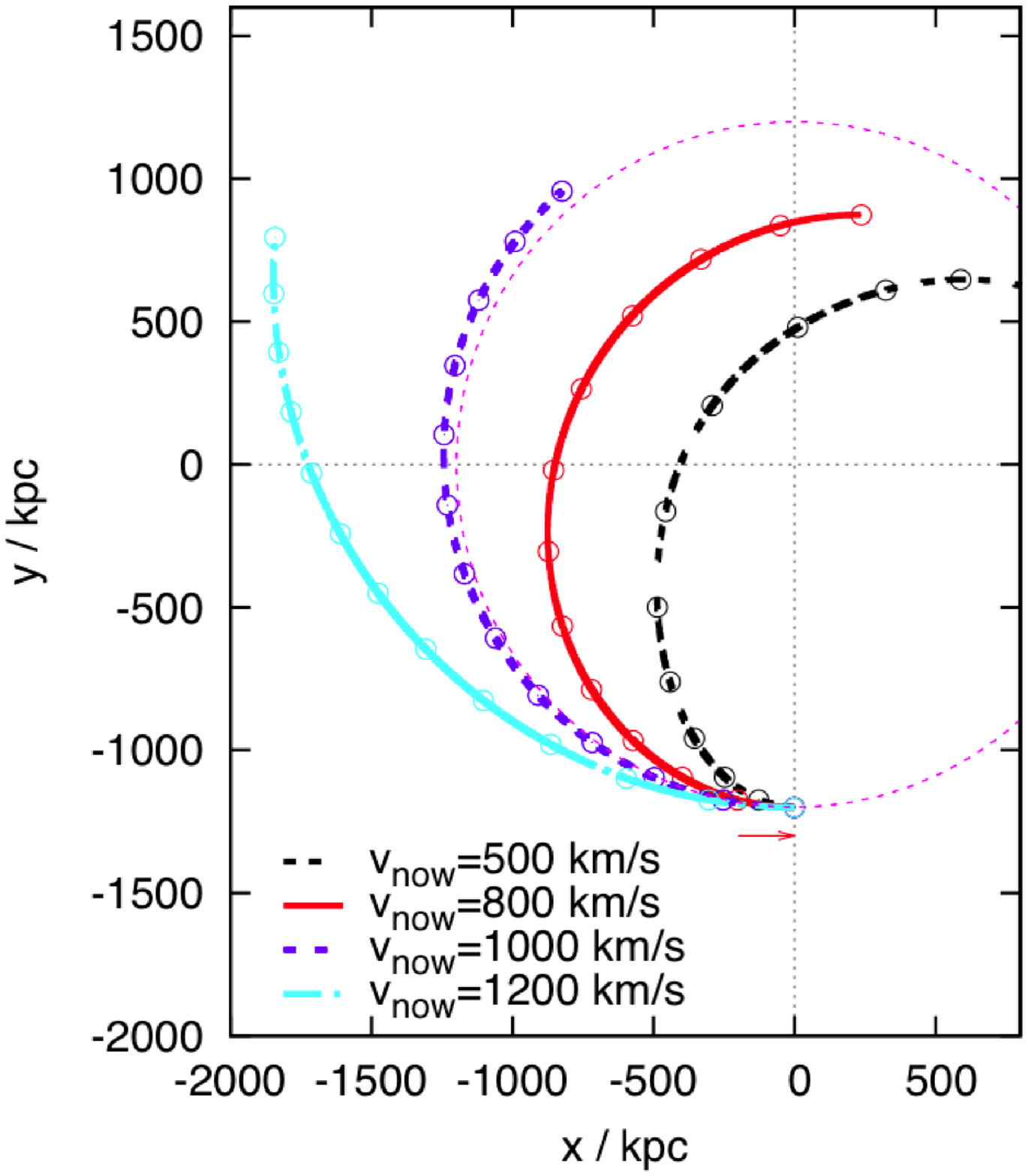}}}
\caption
 {Orbits of a test particle in the potential well of Hydra A moving with different velocities. The red arrow show the position of the observed  LEDA 87445 group at 1.1 Mpc from the center of Hydra A (at coordinates 0,0). The dashed red circle is the Keplerian orbit (i.e. smaller orbits are gravitationally bound with the main cluster). The circles along the orbits are spaced in 0.25 Gyr intervals. }
\label{fig:orbit}
\end{figure}

In Fig. \ref{fig:simul} we show from \cite{roediger15a} an example of a simulated  elliptical galaxy infalling radially in the ICM ambient . A sharp cold front lies at the tip of the structure, with KH rolls on its side. The wake of the infalling structure is composed of a dense region of bound gas that has been pushed back by the ram pressure \citep[dubbed the remnant tail in ][]{roediger15a} and of the actual wake of thin gas stripped through KH instabilities (the stripped wake). The similarity between the observed morphology of our group (Fig. \ref{fig:features}) and the simulated system  in Fig. \ref{fig:simul}  is striking: in addition to the cold front, we also observe the remnant tail of the moving group and the faint, patchy stripped wake.

\cite{vijaya15}, simulating the ram pressure effects on a population of galaxies in isolated cluster environments, found that cluster galaxies are stripped of their gas by the ICM, and that the stripped gas trails the galaxy orbits in the form of wakes before mixing with the ICM.
Under the hypothesis that also the morphology of the infalling LEDA 87445 group, with its bent tail of stripped gas and westward cold front, marks the recent part of the group orbit, we derived a rough estimate of the orbit as follows.
We modelled  the Hydra A cluster as a spherical gravitational potential filled with a hydrostatic ICM. Thus, we could calculate the radial gravitational acceleration in the cluster as a function of radius from the observed ICM density \citep{eckert12}  and temperature profiles \citep{simionescu09}. We then placed a test particle, i.e, a particle that follows only the gravity of the cluster potential, at the current position of LEDA 87445, i.e., 1.1 Mpc south of the cluster centre. Initializing the test particle with the negative of the current velocity of LEDA 87445, $v_{now}$, we could trace back the particleÕs motion in the cluster potential in time. Fig. \ref{fig:orbit} shows the resulting orbits assuming a purely tangential velocity of the group for several magnitudes of $v_{now}$ (the circles along the orbit in the Figure are spaced in 0.25 Gyr intervals). This estimate neglects any deceleration of the group by dynamical friction.

For this cluster model, a current velocity of the test particle such as the one we inferred for the LEDA 87445 group from the stagnation point analysis, $510^{+430}_{-340}$  km/s, would suggest that the group moved on a bound orbit. However if, as discussed in Sect. \ref{sec:motion}, our estimate of the velocity is biased low due to inclination effects,  then the group is  moving on an almost Keplerian orbit and likely making its first passage.

\section{Conclusions}
\label{sec:concl}

In this paper we reported the discovery of the ram-pressure stripped infalling galaxy group in the outskirts of the poor cluster Hydra A/A780.
We have used data from {\it XMM-Newton} and {\it Suzaku} observations, as well as
optical data drawn from the literature, to study the properties of this accreting subgroup.
We summarize our main results as follows:

\begin{enumerate}

\item The X-ray emission peak of the group is at 1.1 Mpc south-east of the Hydra A cluster center. Its X-ray morphology is peculiar as it shows a tail and wake, 760 kpc long (in projection), and, in the opposite direction, a cold front. These features are likely produced by the motion of the infalling group within the ICM.

\item The optical kinematical analysis reveals that the X-ray emission peak coincides with the early-type galaxy LEDA 87445 and that around this galaxy there is an over-density of galaxies, at a c.l. larger that 99\%.

\item The X-ray morphology and optical data, also compared with recent simulations of galaxies infalling in the ICM, give hints on the orbit of the LEDA group leading us to conclude that the group is likely at its first passage onto Hydra A.

\item The gas temperature of the LEDA 87445 group emission peak and along its tail and wake is roughly constant ($\sim 1.2-1.3$ keV), and is lower than the ambient ICM ($\sim 2.4-3.$ keV). This temperature is typical of the virialized plasma of a galaxy group with a mass of a few $10^{13}$ M$_{\odot}$. We estimate that the stripped gas is between 1 and 5$\times 10^{11}$ M$_{\odot}$, i.e. somewhere between half and four fifths of the total observed gas mass.

\item The velocity of the moving group, estimated through the cold front stagnation point argument, is $510_{-430}^{+340}$ km/s relative to the main cluster (corresponding to a Mach number of $0.6_{-0.4}^{+0.5}$) and is comparable to half the sound speed. We find that the gas in the outermost region of the structure was expelled $\sim 1.4-1.6$ Gyr ago.

\item We estimated the timescale required for the gas in the tail to be heated up to the temperature of the ambient ICM and compared this timescale with the age of the structure (i.e., $\sim 1.4-1.6$ Gyr) finding that the suppression factor is at least two orders of magnitude suppressed compared with Spitzer conductivity, in line with the findings for other clusters \citep[e.g.][]{eckert14,ichinohe15}. These long survival times of the stripped gas within the ambient ICM would have important implications on the evolution of the hot gas content in galaxy clusters as the cold gas in the stripped structures could eventually reach the core of the main cluster.

\item Using the relative radial velocity between the galaxy LEDA 87445 and the Hydra A BCG we find that the orbit of the accreting LEDA 87445 group could be inclined with an angle of $\sim 45^\circ$ with respect to the plane of the sky and, therefore, the velocity of the group estimated from the stagnation point analysis would be underestimated due to projection effects. However, more detailed hydrodynamic modelling of this system is required to resolve the full three-dimensional motion of this system.

\end{enumerate}

The striking similarity between the infalling group discovered here and hydrodynamical simulations of similar structures (see Fig. \ref{fig:features} and \ref{fig:simul}) makes this system an ideal target to test our understanding of ram-pressure stripping and ICM physics, e.g. by comparing the morphology of the tail and searching for features like KH instabilities. Deeper, higher resolution observations are however required to perform such a detailed comparison.

\begin{acknowledgements}
We thank Henk Hoekstra and Remco van der Burg for kindly providing us the stacked CFHT image. We acknowledge financial contribution from contract PRIN INAF 2012 (ÓA unique dataset to address the most compelling open questions about
X-ray galaxy clustersÓ). M.G. is supported by NASA through Einstein Postdoctoral Fellowship Award Number PF-160137 issued by the Chandra X-ray Observatory Center, which is operated by the SAO for and on behalf of NASA under contract NAS8-03060. 
This research is based on observations obtained with {\it XMM-Newton}, an ESA science mission with instruments and contributions directly funded by ESA Member States and the USA (NASA), and has made use of data obtained from the {\it Suzaku} satellite, a collaborative mission between the space agencies of Japan (JAXA) and the USA (NASA).
\end{acknowledgements}

\bibliographystyle{aa}
\bibliography{biblio_Jan2016}

\begin{appendix} 
\section{Optical data and kinematical analysis}
\label{app:a1}

We analyzed the \cite{smith04} galaxy sample, that is the largest available homogeneous  sample in the field of view of the Hydra A cluster with galaxies having a measured line-of-sight (LOS) velocity ($V=cz$). The sample contains 42 galaxies and the mean error of this sample is $\delta V\sim 25$ \kss.

\begin{figure}
\centering
\includegraphics[width= 8 cm]{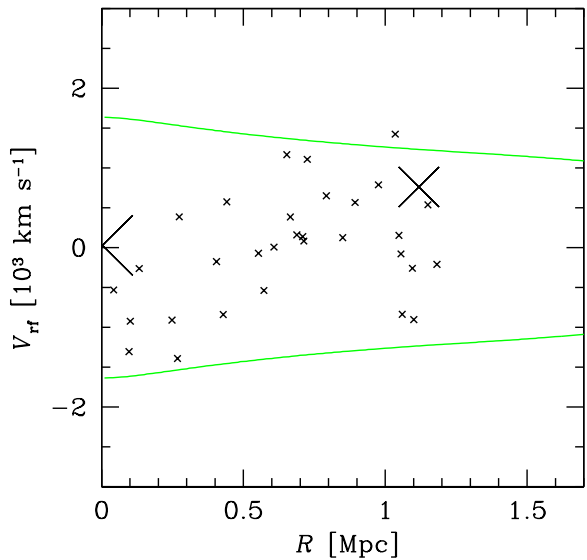}
\caption
{Rest-frame velocity $V_{\rm rf}=(V-<V>)/(1+z)$ vs. projected clustercentric distance $R$ for the 33 cluster members.  Crosses indicate the BGC (center) and LEDA 87445.  Curves contain the region of $V_{\rm rf}$ smaller than the escape velocity. The curves show the limit due to the escape velocity in the cluster.}
\label{figvd}
\end{figure}

For the center of Hydra A, we adopted the position of the brightest cluster galaxy (BCG)  at $V_{\rm BCG}=16\,452$ \ks [R.A.=$09^{\mathrm{h}}18^{\mathrm{m}}05\dotsec69$,  Dec.=$-12\degree05\arcmm 43.8\arcs$ (J2000)].
Following \cite{girardi98} we performed a two step procedure to select cluster members. First, we selected the densest peak in the redshift distribution using the DEDICA reconstruction method \citep{pisani93}. Second, we rejected other possible interlopers in the projected phase-space using the ``shifting gapper'' method \citep{fadda96}. With the first step we detected Hydra A as a peak of 36 galaxies at $z\sim 0.055$ (see Fig.~\ref{fighisto}), whereas the second step led us to reject three more galaxies.
The resulting sample of 33 fiducial members is displayed in Fig.~\ref{figvd} (see also the inset of Fig.~\ref{fighisto} in the Sect. \ref{sec:dynamics}).  By applying the bi-weight estimator \cite{beers90} to the 33 cluster members of the whole system, we obtained a mean cluster redshift of $\left<z\right>=0.0548\pm0.0004$, i.e. $\left<V\right>=(16\,421\pm121)$ \kss.

We computed the LOS velocity dispersion, $\sigma_{V}$, by using the bi-weight estimator and applying the cosmological correction and the standard correction for velocity errors \citep{danese80}.
We obtained $\sigma_V=723_{-72}^{+93}$ \kss, where errors are estimated through a bootstrap technique. This estimate is within the range of a few alternative estimates presented by \cite{durret09} and with that expected for clusters with X-ray temperatures kT$\sim 3.5$ keV \citep{sato12}, in the assumption of the density-energy equipartition between ICM and galaxies.  In fact, we estimated $\beta_{\rm spec}\sim 1.05$, where $\beta_{\rm   spec}=\sigma_V^2/(kT/\mu m_{\rm p})$ with $\mu=0.58$ the mean molecular weight and $m_{\rm p}$ the proton mass, and $\beta_{\rm  spec}=1$ is expected in the case of density-energy equipartition. Based on the value of $\sigma_V$ we estimated a mass $M_{200}\sim 4$ \mqua within a radius $R200_{opt}\sim 1.5$ \h \citep{munari13}.

The bright galaxy LEDA 87445 (R.A.=$09^{\mathrm{h}}18^{\mathrm{m}}27\dotsec47$, Dec.=$-12\degree22\arcmm 23.9\arcs$ (J2000)) is a cluster member lying at the high velocity part of the cluster-galaxies velocity distribution ($V_{\rm LEDA}=17\,224$ \kss).  Inspection of Fig.~\ref{figvd} suggests that regions around the BCG and LEDA 87445 are populated by galaxies with different kinematics.
To investigate this point we applied the Dressler \& Schectman test \citep[; DS-test]{dressler88}. This is a powerful test for 3D substructure, which is sensitive in samples down to 30 member galaxies, at least in the case of important mergers \citep{pinkney96}.  It analyzes the deviations of the local kinematics, as based on the i-th galaxy and its 10 neighbors, from the global kinematics. The DS-test kinematical estimator is based on both mean velocity and velocity dispersion.
We refer to the original paper for details as well as to \cite{barrena11} where two alternative kinematical estimators are presented. These estimators separately consider the contributions to the mean velocity and velocity dispersion (DS$\left<V\right>$ and DS$\sigma_{V}$-tests).  We checked the significance of the existence of substructure by running 1000 Monte Carlo simulations, randomly shuffling the galaxy velocities.  We detected a substructure with a c.l. of 99.4\% (and 99.7\% in the DS$\left<V\right>$ case). Similar results ( 99.1\% and 99.7\%) are obtained using local groups of 6 galaxies, corresponding to $N_{\rm local}=\sqrt{N_{\rm gal}}$ as suggested by \cite{bird94} (see Figure ~\ref{figdssegno5v} in Sect. \ref{sec:dynamics}).

Following \cite{girardi96} we also searched for the presence of a velocity gradient by performing a multiple linear regression fit to the observed velocities with respect to the galaxy positions in the plane of the sky.  The significance is again based on 1000 Monte Carlo simulated clusters. We found only a marginal evidence (at 90\% c.l.) of the existence of a velocity gradient.  The position angle $PA=143_{-21}^{+29}$ (measured counter--clock--wise from north) has a large error but agrees with Fig.~\ref{figdssegno5v} in detecting higher velocities in the S-SE region of Hydra A.

\begin{figure}
\centering
\includegraphics[width=8cm]{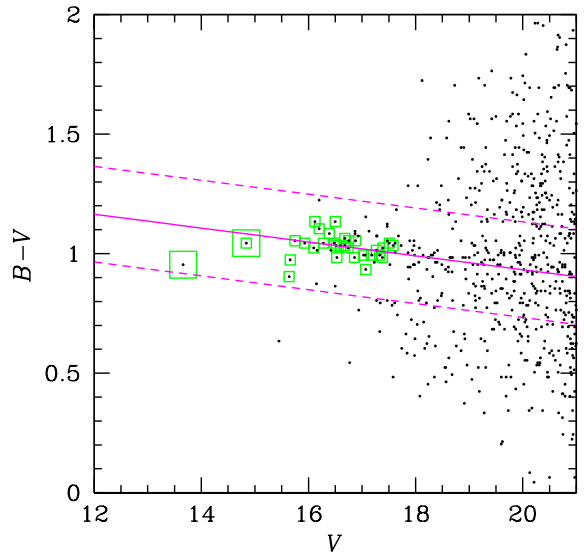}
\caption
{$B-V$ vs. $V$ diagram for galaxies in the  photometric WINGS catalog having $V<21$. Green squares indicate spectroscopic  members and, in particular, the two large squares indicate the BGC and LEDA 87445.  The solid line gives the best-fit CMR as determined on  spectroscopic member galaxies; the dashed lines define the region within which photometric cluster members are selected.  }
\label{figcm}
\end{figure}

We analyzed also the spatial distribution of the 33 spectroscopic member galaxies by using the 2D adaptive-kernel method of \cite{pisani96} (hereafter 2D-DEDICA). In addition to the central main density peak, we also detected a SSE peak coincident to LEDA 87445, but only significant at the $98\%$ c.l.. In the study of the 2D galaxy distribution the use of the spectroscopic sample is limited by both the poor number of galaxies and the possible incompleteness due to unavoidable constraints in the design of the fiber positioning during the spectroscopy.  The analysis of the published WINGS\footnote{http://web.oapd.inaf.it/wings/} photometric catalog \citep{varela09} offers an additional description of the 2D cluster structure.  The WINGS catalog has been based on deep ($B$,$V$) wide field images ($\sim$35\arcmin$\times$35\arcmin). It is 90\% complete at $V\sim$21.7 and the star/galaxy classification of objects $V<20$ was checked visually.

\begin{figure}
\centering
\includegraphics[width=8cm]{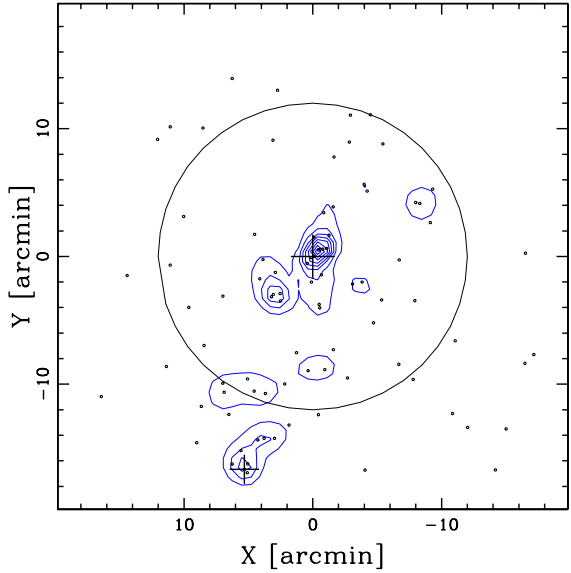}
\caption{Spatial distribution on the sky of the 89 photometric likely  cluster members with $V\le 18.5$ (black small points). The contour  map is obtained with the 2D-DEDICA method (blue lines).  The larger  and smaller crosses indicate the position of BCG and LEDA 87445.  The plot  is centered on the cluster center and the circle contains the region  within a radius equal to 12\arcmm, $\sim 0.5 R200_{opt}$.}
\label{figk2}
\end{figure}

In this study we focus on the red sequence galaxies, which are well known good tracers of the cluster substructure \citep[e.g.][]{lubin00} and above all are useful to minimize the field contamination.
We retrieved from CDS Vizier photometric data for 1632 WINGS galaxies in the Hydra A field. We corrected the Galactic extinction $A_B=0.148$ and $A_V=0.112$ following values listed by NED\footnote{NASA/IPAC Extragalactic Database which is operated by the Jet Propulsion  Laboratory, California Institute of Technology, under contract with the NASA.}
We selected photometric members on the basis of ($B-V$ vs. $V$) color-magnitude relation (CMR), in particular we used aperture magnitudes [MAG(10kpc)] for $B-V$ and total magnitude [MAG\_AUTO] for $V$.  To determine the CMR we applied a recursive 2$\sigma$-clipping fitting procedure to the spectroscopic member galaxies. During this procedure, since these members sample a small magnitude range, we fixed the slope equal to $-0.047$, which is the median slope of WINGS clusters \citep{valentinuzzi11}. We selected a subsample of 24 members for which we fit the OLS($y/x$) relation $B-V=1.513(\pm 0.171)-0.029(\pm 0.010)\times V$, the slope is consistent with that obtained for colors estimated in 5 kpc \citep{valentinuzzi11}. In agreement with \cite{valentinuzzi11} we considered only galaxies within a color range of 0.2 (see Fig.~\ref{figcm}). Moreover, we limited our analysis to galaxies with $V\le 18.5$ (89 in the whole photometric catalog), i.e. $\sim 3$ mag after $M^*$, to reduce the contamination of non member galaxies.  Figure~\ref{figk2} shows the contour map where we detected three density peaks significant at the $99\%$ c.l., among which a SSE peak at the location of the LEDA 87445 galaxy. This peak has a relative density $\rho_S=0.4$ with respect to the central main peak, although the number of related members is very small (5 galaxies).

\end{appendix}

\end{document}